\documentclass[12pt]{article}
\usepackage[a4paper,text={16.8cm,22.4cm}]{geometry}
\usepackage{amsmath,amsfonts,braket,slashed,amssymb,tikz,bm,psfrag,graphicx,color,dsfont,euscript}
\usepackage[small,labelfont=bf]{caption}

\RequirePackage[sort&compress,square,comma,numbers]{natbib}
\allowdisplaybreaks
\begin{document}

\begin{titlepage}

\begin{flushright}
MZ-TH/19-056\\[1mm]
August 29, 2019
\end{flushright}

\vspace{0.5cm}
\begin{center}
\Large\bf
Infrared Singularities of Scattering Amplitudes and N$^3$LL Resummation for $\bm{n}$-Jet Processes
\end{center}

\vspace{0.5cm}
\begin{center}
Thomas Becher\,$^a$ and Matthias Neubert\,$^{b,c}$\\
\vspace{0.4cm}
{\sl $^a$\,Albert Einstein Center for Fundamental Physics, Institut f\"ur Theoretische Physik\\ 
Universit\"at Bern, Sidlerstrasse 5, CH-3012 Bern, Switzerland\\[2mm]
$^b$\,PRISMA$^+$ Cluster of Excellence {\em\&} Mainz Institute for Theoretical Physics\\ 
Johannes Gutenberg University, 55099 Mainz, Germany\\[2mm]
$^c$\,Department of Physics {\em\&} LEPP, Cornell University, Ithaca, NY 14853, U.S.A.}
\end{center}

\vspace{0.2cm}
\begin{abstract}\noindent
We revisit the multi-loop structure of the anomalous-dimension matrix governing the infrared divergences of massless $n$-particle scattering amplitudes in non-abelian gauge theories. In particular, we derive its most general form at four-loop order, significantly simplifying corresponding expressions given previously. By carefully reevaluating the constraints imposed by two-particle collinear limits, we find that at four-loop order color structures involving $d_R^{abcd}$, the symmetrized trace of four group generators, appear along with cusp logarithms $\ln[\mu^2/(-s_{ij})]$. As a consequence, naive Casimir scaling of the cusp anomalous dimensions associated with the quark and gluon form factors is violated, while a generalized form of Casimir scaling still holds. Our results provide an important ingredient for resummations of large logarithms in $n$-jet cross sections with next-to-next-to-next-to leading logarithmic (N$^3$LL) accuracy.
\end{abstract}
\vfil

\end{titlepage}

\tableofcontents

\newpage
\section{Introduction}

Understanding the structure of infrared (IR) singularities of gauge-theory scattering amplitudes is an important problem. On one hand, this helps in unveiling the deeper structure of quantum field theory in higher orders of perturbation theory. On the other, it also has many practical applications. In particular, the ability to predict the IR singularities of $n$-particle amplitudes enables one to systematically resum large logarithmic corrections to cross sections and differential distributions for many important collider processes, leading to a higher precision in the calculation of these observables. 

The problem of predicting the structure of IR singularities of on-shell $n$-particle scattering amplitudes in massless QCD simplifies, if one realizes that they can be put in one-to-one correspondence with ultraviolet (UV) divergences of operators defined in soft-collinear effective theory (SCET) \cite{Becher:2009cu}. This relation implies that IR divergences can be studied by means of standard renormalization-group techniques -- a concept that had been developed earlier in the context of theories of Wilson lines \cite{Korchemskaya:1994qp}. The IR divergences of $n$-point scattering amplitudes can be absorbed into a multiplicative renormalization factor $\bm{Z}$, which can be derived from an anomalous dimension $\bm{\Gamma}$. Both objects are matrices in color space, i.e.\ they mix amplitudes with the same particle content but different color structures. The predictive power of this approach relies on the fact that the anomalous dimension is tightly constrained by the structure of the effective field theory: soft-collinear factorization implies that it is given by the sum of a soft and a collinear contribution, 
\begin{equation}\label{softcollfact}
   \bm{\Gamma}(\{\underline{s}\},\mu) 
   = \bm{\Gamma}_s(\{\underline{\beta}\},\mu) + \sum_{i=1}^n\,\Gamma_c^i(L_i,\mu)\,\bm{1} \,,
\end{equation}
and given that there are no interactions among different collinear sectors of SCET \cite{Bauer:2000yr,Bauer:2001yt,Bauer:2002nz,Beneke:2002ph}, all non-trivial color and momentum dependence is encoded in the soft anomalous dimension $\bm{\Gamma}_s$. 

The total anomalous dimension $\bm{\Gamma}$ depends on the $n(n-1)/2$ kinematical variables $s_{ij}\equiv 2\sigma_{ij}\,p_i\cdot p_j+i0$, where the sign factor $\sigma_{ij}=+1$ if the momenta $p_i$ and $p_j$ are both incoming or outgoing, and $\sigma_{ij}=-1$ otherwise. We denote the collection of these variables by $\{\underline{s}\}$. It also depends on the color generators $\bm{T}_i$ of the $n$ particles. We suppress this dependence in the argument of the anomalous dimension but indicate it by the use of the boldface symbol $\bm{\Gamma}$, which shows that the anomalous dimension is a matrix in color space. We use the color-space formalism, in which amplitudes are treated as $n$-dimensional vectors in color space \cite{Catani:1996vz}. $\bm{T}_i$ is the color generator associated with the $i^{\rm th}$ particle in the scattering amplitude, which acts as an $SU(N_c)$ matrix on the color indices of that particle. 

The soft anomalous dimension $\bm{\Gamma}_s$ is the anomalous dimension of an operator built out of $n$ soft Wilson lines, one for each external particle, directed along the momentum of that particle and defined in the appropriate representation of $SU(N_c)$. The dependence of the soft anomalous dimension on the external momenta $p_i$ of the particles is encoded via so-called cusp angles $\beta_{ij}$ (with $i\ne j$), which for slightly off-shell, massless particles are defined as
\begin{equation}
   \beta_{ij} = \ln\frac{(-s_{ij})\,\mu^2}{(-p_i^2-i0)(-p_j^2-i0)} 
   = L_i + L_j - \ln\frac{\mu^2}{-s_{ij}} \,.
\end{equation}
The collinear anomalous dimensions $\Gamma_c^i$ are single-particle terms, which are diagonal in color space and each depend on a single collinear scale $L_i=\ln[\mu^2/(-p_i^2-i0)]$. To all orders in perturbation theory, they have the form \cite{Becher:2003kh}
\begin{equation}\label{gammaci}
   \Gamma_c^i(L_i,\mu)
   = - \Gamma_{\rm cusp}^i(\alpha_s)\,L_i + \gamma_c^i(\alpha_s) \,,
\end{equation}
where the coefficients $\Gamma_{\rm cusp}^i(\alpha_s)$ is called the cusp anomalous dimension of particle $i$ \cite{Korchemskaya:1992je}. The fact that the total anomalous dimension must be independent of the collinear scales $p_i^2$ when we combine the soft and collinear contributions implies the differential equation \cite{Gardi:2009qi,Becher:2009qa}
\begin{equation}\label{ourconstraint}
   \frac{d\bm{\Gamma}_s(\{\underline{\beta}\},\mu)}{dL_i}
   = \sum_{j\ne i}\,\frac{\partial\bm{\Gamma}_s(\{\underline{\beta}\},\mu)}{\partial\beta_{ij}}
   = \Gamma_{\rm cusp}^i(\alpha_s)\,\bm{1} \,,
\end{equation}
where the expression on the right-hand side is a unit matrix in color space. 

This relation provides an important constraint on the momentum and color structures that can appear in the soft anomalous-dimension matrix. Because the kinematical invariants $s_{ij}$ can be assumed to be linearly independent, relation (\ref{ourconstraint}) implies that $\bm{\Gamma}_s$ depends only linearly on the individual cusp angles $\beta_{ij}$. The only exception would be a more complicated dependence on combinations of cusp angles, in which the collinear logarithms cancel. The simplest such combination is
\begin{equation}\label{CCR}
   \beta_{ijkl} = \beta_{ij} + \beta_{kl} - \beta_{ik} - \beta_{jl} 
   = \ln\frac{(-s_{ij})(-s_{kl})}{(-s_{ik})(-s_{jl})} \,,
\end{equation} 
which coincides with the logarithm of the conformal cross ratio $\rho_{ijkl}$ defined in \cite{Gardi:2009qi}. For simplicity, we will use the term ``conformal cross ratio'' in the following also when referring to $\beta_{ijkl}$. This quantity obeys the symmetry properties
\begin{equation}\label{betasym}
   \beta_{ijkl} = \beta_{jilk} = - \beta_{ikjl} = - \beta_{ljki} 
   = \beta_{klij} \,.
\end{equation}
It is easy to show that any combination of cusp angles that is independent of collinear logarithms can be expressed via such cross ratios. Moreover, given four particle momenta there exist only two linearly independent conformal cross ratios, since
\begin{equation}\label{betazerosum}
   \beta_{ijkl} + \beta_{iklj} + \beta_{iljk} = 0 \,,  
\end{equation}
and all other index permutations can be obtained using the symmetry properties in (\ref{betasym}). Any function of conformal cross ratios provides a solution to the homogeneous differential equation associated with (\ref{ourconstraint}), and hence it can always be added to any particular solution of the equation.

Another powerful constraint arises from the non-abelian exponentiation theorem \cite{Gatheral:1983cz,Frenkel:1984pz}, which implies that only the color structures associated with fully connected gluon webs, whose ends can be attached in arbitrary ways to the $n$ Wilson lines, contribute to the soft anomalous dimension \cite{Gardi:2009qi,Becher:2009qa}. This severely restricts the color structures that can arise in higher orders of the loop expansion. The generalization of the concept of ``webs'' to multi-particle amplitudes has been discussed in detail in \cite{Gardi:2010rn,Gardi:2013ita}.

Up to two-loop order, the constraints mentioned above imply that a simple dipole formula describes the anomalous dimension for arbitrary scattering processes of $n$ massless particles \cite{Becher:2009cu,Gardi:2009qi,Becher:2009qa}, in accordance with explicit calculations \cite{Aybat:2006wq,Aybat:2006mz}. A more complicated formula describes processes in which some or all of the participating particles are massive \cite{Becher:2009kw,Ferroglia:2009ep,Ferroglia:2009ii}. Here we reconsider the case of massless particles, where starting from three-loop order non-trivial correlations between three or more particles arise \cite{Ahrens:2012qz}. The explicit structure of the three-loop three- and four-particle correlations was derived in \cite{Almelid:2015jia}. 

The functional form of the multi-particle correlations and their dependence on the kinematic variables $s_{ij}$ and $\beta_{ijkl}$ is further constrained by collinear factorization \cite{Becher:2009qa}. When two particles in either the initial or the final state of a scattering process become collinear, an $n$-particle scattering amplitude splits into an $(n-1)$-particle amplitude times a process-independent splitting amplitude ${\bf Sp}(\{p_1,p_2\},\mu)$, which involves the momenta and color generators of the collinear particles only \cite{Berends:1988zn,Mangano:1990by,Bern:1995ix,Kosower:1999xi}. The fact that the anomalous dimension of the splitting amplitude defined as
\begin{equation}
   \frac{d}{d\ln\mu}\,\mbox{\bf Sp}(\{p_1,p_2\},\mu)
   = \bm{\Gamma}_{\rm Sp}(\{p_1,p_2\},\mu)\,\mbox{\bf Sp}(\{p_1,p_2\},\mu) \,,
\end{equation} 
must be independent of the momenta and color generators of the particles not involved in the splitting process yields the  non-trivial constraint \cite{Becher:2009qa}
\begin{equation}\label{collconstraint}
   \bm{\Gamma}_{\rm Sp}(\{p_1,p_2\},\mu) 
   = \bm{\Gamma}(\{p_1,\dots,p_n\},\mu) - \bm{\Gamma}(\{P,p_3\dots,p_n\},\mu) \big|_{\bm{T}_P\to\bm{T}_1+\bm{T}_2} \,,
\end{equation}
where also the right-hand side must be independent of the momenta $p_3,\dots,p_n$ and the corresponding color generators. Collinear factorization is known to be violated in the space-like region, when one of the collinear particles is in the initial state while the other belongs to the final state \cite{Catani:2011st,Forshaw:2012bi}. For our purposes, however, we can always assume that the two collinear particles 1 and 2 both belong to the final state. The high-energy (``Regge'') limit imposes an interesting additional constraint on $n$-particle scattering amplitudes \cite{Bret:2011xm,DelDuca:2011ae}. The point is that the leading IR singularities of the Regge slopes are correctly described by the dipole conjecture, so extra contributions from multi-particle correlation terms must only give rise to subleading logarithms. 

In this paper we revisit our previous analyses \cite{Becher:2009qa,Ahrens:2012qz} of the structure of the anomalous-dimension matrix $\bm{\Gamma}$ for $n$-particles scattering amplitudes in massless Yang-Mills theory. We begin with some comments on the workings of non-abelian exponentiation and the definition of connected webs for $n$-particle amplitudes. We then show how these webs can be decomposed into color structures that are symmetrized with respect to the external particle indices. Our master formula for the anomalous dimension $\bm{\Gamma}$, which has been simplified compared to earlier expressions due to the fact that we have unravelled some new color identities, is presented in relation (\ref{magic}) in Section~\ref{sec:4}, where we also summarize the present knowledge of the various perturbative coefficient functions entering the result. The constraints on the coefficient functions implied by the proper factorization in two-particle collinear limits are derived in Section~\ref{sec:5}. Two interesting phenomenological consequences of our results are discussed in Section~\ref{sec:6}, where we quote the anomalous dimension relevant for the resummation of large logarithms in collider cross sections at N$^3$LL order as well as the anomalous dimensions governing the IR singularities of three-particle scattering amplitudes.

\section{Non-abelian exponentiation and connected webs}

\begin{figure}
\begin{center}
\begin{tabular}{ccccc}
\psfrag{h}[]{$\cal{M}$}
\includegraphics[width=0.22\textwidth]{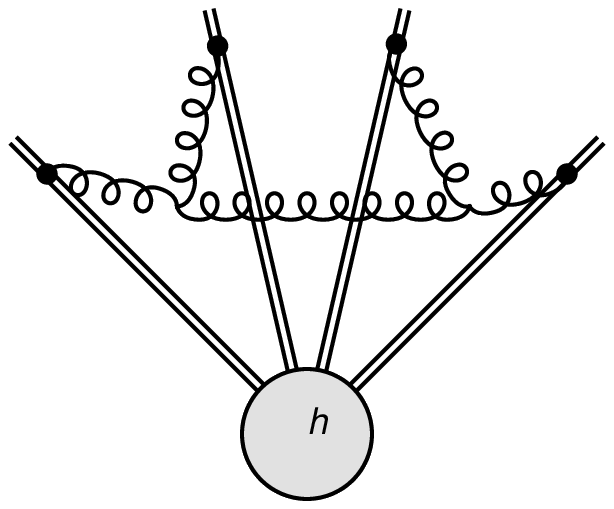} &&
\psfrag{h}[]{$\cal{M}$}
\includegraphics[width=0.22\textwidth]{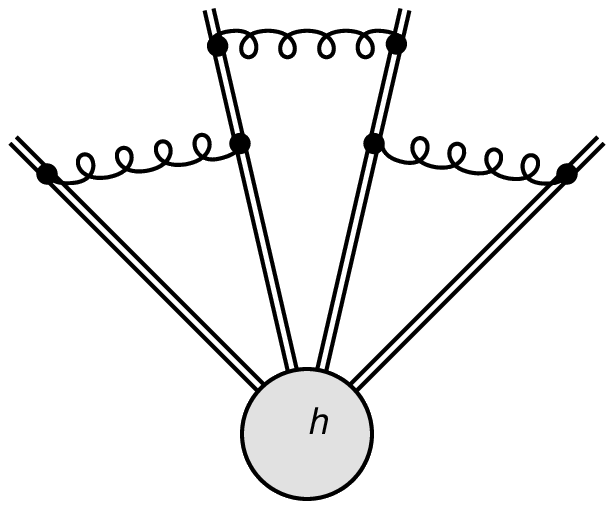} &&
\psfrag{h}[]{$\cal{M}$}
\includegraphics[width=0.22\textwidth]{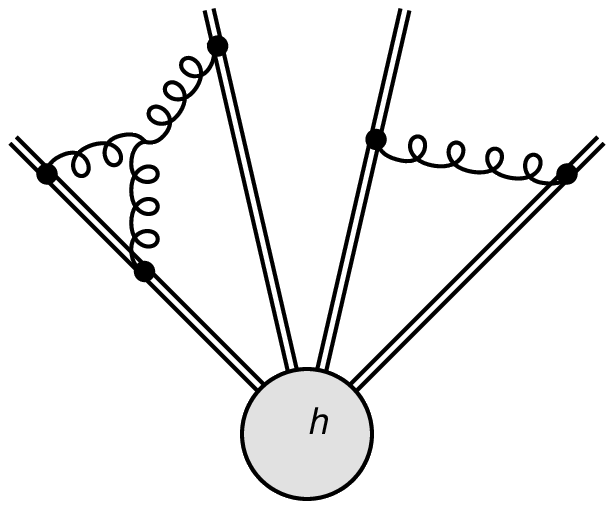} 
\end{tabular}
\vspace{2mm}
\caption{\label{fig:colorConnected} 
Representative three-loop diagrams contributing to the soft function $S$ associated with a four-particle scattering amplitude. The light-like soft Wilson lines are represented by double lines and multiply the hard amplitude $\cal{M}$ indicated by the gray blob. The left diagram is fully connected and therefore also color connected. The middle diagram is connected but not fully connected. It has a color-connected part with the same color structure as the diagram on the left. The right diagram is disconnected.}
\end{center}
\end{figure}

Since the color structure of the collinear anomalous dimension is trivial, the hard anomalous dimension inherits the color structures of the soft anomalous dimension $\bm{\Gamma}_s(\{\underline{\beta}\},\mu)$ in \eqref{softcollfact}. The soft anomalous dimension governs the ultraviolet (UV) poles of a soft function $S$, which is given by a matrix element of a product of soft light-like Wilson lines in the directions of the external particles. Figure~\ref{fig:colorConnected} shows a few representative Feynman graphs contributing to $S$ at ${\cal O}(\alpha_s^3)$ in perturbation theory. The soft anomalous dimension is derived from the coefficient of the $1/\epsilon$ pole in the exponent $\widetilde{S}$ defined through $S=\exp(\widetilde{S})$.

\begin{figure}
\begin{center}
\includegraphics[width=0.5\textwidth]{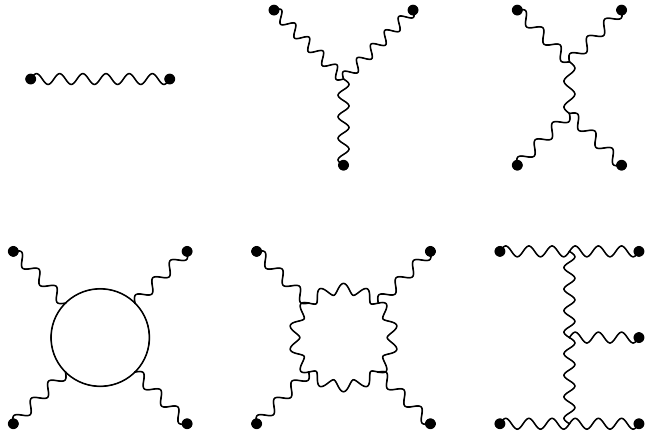}
\vspace{4mm}
\caption{\label{fig:webs}
Color-connected webs appearing up to four-loop order in the soft anomalous dimension $\bm{\Gamma}_s$. The webs represented by these graphs are the color structures that arise if the wavy lines are replaced by gluons in the corresponding (fully connected) tree-level Feynman graphs.}
\end{center}
\end{figure}

The higher-order corrections to the soft function are severely constrained. In fact, in an abelian theory (with massive fermions), soft Wilson-line matrix elements are almost trivial, since the higher-order contributions are obtained by exponentiating the one-loop result, and hence $\widetilde{S}$ is saturated at one-loop order. This simple exponentiation does not hold in non-abelian theories, but the higher-order corrections to the exponent only arise from a restricted set of color structures, as first demonstrated by Gatheral \cite{Gatheral:1983cz}. The color structures arising up to four-loop order are shown in Figure~\ref{fig:webs}. They were called ``color-connected webs'' by Frenkel and Taylor \cite{Frenkel:1984pz}. In the following, it will be important to distinguish the terms ``fully connected'' and ``color connected''. The exponent $\widetilde{S}$ also gets contributions from diagrams, in which the gluons are not directly connected with each other, but whose color structure is equal to the color structure of a fully connected diagram after using the group identity $[\bm{T}^a,\bm{T}^b]=if^{abc}\,\bm{T}^c$ to ``connect'' two gluons. A diagram is called ``fully connected'', if it stays connected when one cuts Wilson-line propagators. The first graph in Figure~\ref{fig:colorConnected} shows an example. By definition, a fully connected diagram is also color connected, but also diagrams which are not fully connected can contain color-connected pieces. An example is shown by the second graph in the figure. Only disconnected diagrams such as the third one cannot give rise to color-connected contributions.

The original papers \cite{Gatheral:1983cz,Frenkel:1984pz} on non-abelian exponentiation were focussing on the form-factor case, which involves soft emissions from only two Wilson lines. The generalization to multiple Wilson lines has been developed in \cite{Gardi:2010rn,Mitov:2010rp,Gardi:2011wa,Gardi:2011yz,Gardi:2013ita,Vladimirov:2014wga,Vladimirov:2015fea}. It is based on an efficient method to evaluate the diagrammatic contributions to the exponent $\widetilde{S}$ introduced in \cite{Laenen:2008gt,Gardi:2010rn}. The technique is called the ``replica trick'' and is well known in statistical physics (see e.g.\ \cite{spinGlass}), where it can be used to compute the logarithm of the partition function. It is based on the identity
\begin{equation}
   \widetilde{S} = \ln S = \lim_{N\to 0} \frac{S^N-1}{N} \,.
\end{equation}
The trick consists in evaluating $S^N$ with $N$ replicas of QCD. The contribution to the exponent $\widetilde{S}$ is then obtained after expanding the result for $S^N$ in a Taylor series in $N$ and picking up the linear term. To get the $N^{\rm th}$ power of $S$, one has to order the color matrices of the different replicas on the Wilson line, i.e.\ one starts with the color matrices associated with the first copy and ends with the ones of the $N^{\rm th}$ copy when moving along the Wilson line. 

An efficient way to compute the diagrams of the replicated theory is to draw the usual (non-replicated) QCD Wilson-line diagrams and then assign different replicas to different gluons in the diagram. To get the result in the replicated theory, one then has to add the proper combinatorial factor for each replica assignment. For example, if the diagram is fully connected, only a single replica can contribute, because the different replicas are independent copies of QCD and do not interact with each other. Since there are $N$ replicas, the combinatorial factor is $N$ and the diagram directly contributes to $\widetilde{S}$. This gives the basic, but important statement that fully connected diagrams contribute to the exponent $\widetilde{S}$. Given that these diagrams are color connected, it is clear that the structures shown in Figure~\ref{fig:webs} are indeed present in $\tilde{S}$. What remains to be shown is that the exponent does not involve any color-disconnected contributions from other diagrams.

It is easy to show that disconnected diagrams do not give a contribution to the exponent, since they scale as $N^2$, as each part of the diagram can involve a different replica. The interesting class of diagrams, which we will study in the following, are connected diagrams which become disconnected by cutting one or more Wilson lines, i.e.\ diagrams which are connected but not fully connected. For such diagrams the appropriate combinatorial factor for a contribution with $M$ different replicas is
\begin{equation}\label{comb}
   \frac{N!}{M!\,(N-M)!} = \frac{(-1)^{M-1}}{M}\,N + {\cal O}(N^2) \,.
\end{equation}
There are in general $M!$ factorial possibilities to order the replicas in the diagram. For example, in a diagram in which cutting Wilson lines leads to two disconnected pieces, one can assign two different replicas $I$ and $J$, but we can have $I<J$ or $J<I$, each of which contributes according to (\ref{comb})  with a factor $-1/2$ to the exponent $\widetilde{S}$. 

\begin{figure}
\begin{center}
\begin{tabular}{ccc}
\psfrag{a}[]{$\vdots$}\psfrag{b}[]{$\vdots$}
\psfrag{c}[]{$\bm{C}$}\psfrag{d}[]{$\bm{D}$}
\includegraphics[width=0.35\textwidth]{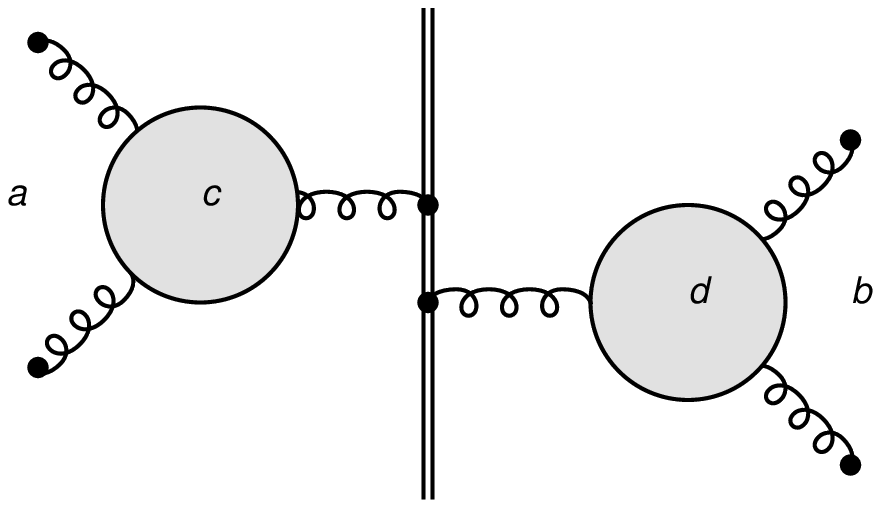} &&
\psfrag{a}[]{$\vdots$}\psfrag{b}[]{$\vdots$}
\psfrag{c}[]{$\bm{E}$}\psfrag{d}[]{$\bm{D}$}
 \includegraphics[width=0.35\textwidth]{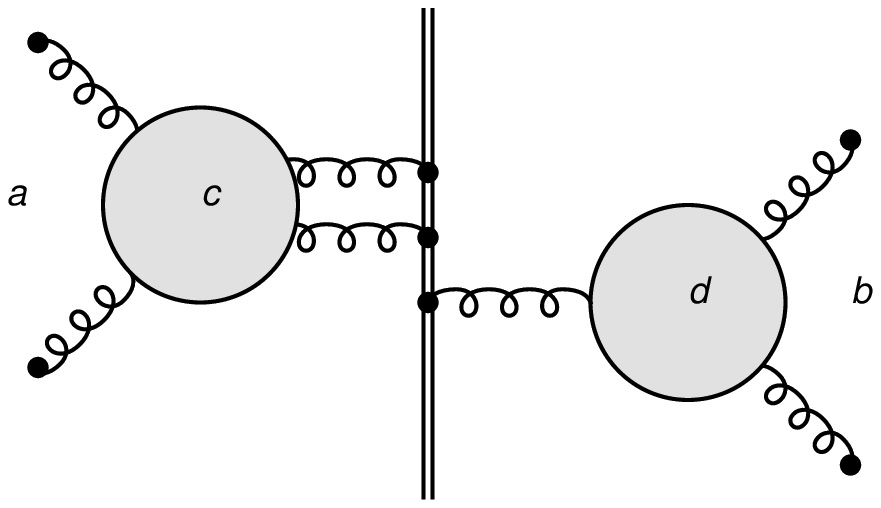}  
\end{tabular}
\vspace{2mm}
\caption{\label{fig:conn} Two examples of gluon clusters connecting to a Wilson line.}
\end{center}
\end{figure}

Let us evaluate one example in detail, namely a contribution with two disjoint connected gluon clusters attaching to a single Wilson line at leg $i$. The corresponding type of diagram is depicted on the left-hand side of Figure~\ref{fig:conn} and has the form
\begin{equation}
   D = F\,\bm{C}^a\!\bm{D}^b\,\bm{T}_i^a \bm{T}_i^b \,.
\end{equation} 
The function $F$ contains the kinematic information of the diagrams, while $\bm{C}^a$ and $\bm{D}^b$ account for the color structures of the two connected clusters. The factors $\bm{C}^a$ and $\bm{D}^b$ are functions of the color generators of other lines, but since the clusters are disjoint and themselves fully connected, they will be the same in the replicated theory. If we assign replica numbers $I$ and $J$ to the two clusters, there are three cases to consider: $I=J$, $I<J$, and $I>J$. The associated contributions to the exponent are as follows:
\begin{equation}\label{eq:repl}
\begin{aligned}
   &I=J: \quad & &F\,\bm{C}^a\!\bm{D}^b\,\bm{T}_i^a \bm{T}_i^b \,, \\
   &I<J: &-\frac{1}{2}\, &F\,\bm{C}^a\!\bm{D}^b\,\bm{T}_i^a \bm{T}_i^b \,, \\[-1mm]
   &I>J: &-\frac{1}{2}\, &F\,\bm{C}^a\!\bm{D}^b\,\bm{T}_i^b \bm{T}_i^a \,.
\end{aligned}
\end{equation}
Note that the color matrices in the third contribution were replica-ordered, i.e.\ reordered so that the replica-number increases along the Wilson line. Summing up the three terms, one obtains for the contribution of the diagram to the exponent $\widetilde{S}$ 
\begin{equation}
   \widetilde{D} = \frac{1}{2}\,F\,\bm{C}^a\!\bm{D}^b\,[\bm{T}_i^a,\bm{T}_i^b]  
   = \frac{i}{2}\,F f^{abc}\,\bm{C}^a\!\bm{D}^b\,\bm{T}_i^c \,,
\end{equation}
which has a fully-connected color structure, as expected. Of course, we could have split up the original diagram into a color-connected piece and a remainder by rewriting
\begin{equation}
   \bm{T}_i^a \bm{T}_i^b  = \frac{1}{2}\,[\bm{T}_i^a,\bm{T}_i^b] + \frac{1}{2}\,\{ \bm{T}_i^a,\bm{T}_i^b \} \,.
\end{equation}
The replica treatment eliminates the contribution of the symmetric, color-disconnected piece to the exponent. More generally, the replica method acts in the space of color structures related to each other by permuting color generators on the Wilson lines. Within this space, it acts as a projection onto the structures in the exponent. In \cite{Gardi:2010rn} the map onto color structures in the exponent $\widetilde S$ was written in matrix form, and one finds that the corresponding mixing matrices $R$ are indeed projections, i.e.\ $R^2=R$. In our trivial example above, the mixing matrix reads
\begin{equation}
   R = \frac{1}{2}\,\bigg( \begin{matrix} \phantom{-}1 & -1~ \\
                           -1 & \phantom{-}1~ \end{matrix} \bigg)
\end{equation}
and acts on the two color structures $\bm{T}_i^a \bm{T}_i^b$ and $\bm{T}_i^b \bm{T}_i^a$. Many explicit examples of such matrices were provided in \cite{Gardi:2010rn}.

The result of our computation (\ref{eq:repl}) can be compactly summarized as a substitution rule
\begin{equation}
   \bm{T}_i^a \bm{T}_i^b \to \frac{i}{2}\,f^{abc}\,\bm{T}^c \,.
\end{equation}
The analogous result for attaching three different clusters to a single Wilson line reads
\begin{equation}
   \bm{T}_i^a \bm{T}_i^b \bm{T}_i^c \to \frac{1}{6} \left( 2 f^{ade} f^{bce}- f^{ace} f^{bde} \right) \bm{T}_i^d \,.
\end{equation}
This color structure consists of two contracted structure constants, i.e.\ two instances of the third color structure in Figure~\ref{fig:webs}. Repeating the exercise with four gluons, the maximum number which can arise at four-loop order, we obtain a linear combination of terms with three connected $f^{abc}$ symbols, corresponding to the last color structure in Figure~\ref{fig:webs}. 

Let us consider a more interesting example, in which two lines of a connected gluon cluster are attached to the same Wilson line, as depicted on the right-hand side of Figure~\ref{fig:conn}. This gives rise to the color structure $\bm{E}^{ab}\bm{D}^c\,\bm{T}_i^a\bm{T}_i^b\bm{T}_i^c$, where $a$ and $b$ connect to the same cluster and thus must be part of the same replica. In analogy with \eqref{eq:repl}, the replica trick leads to
\begin{equation}
   \bm{T}_i^a\bm{T}_i^b\bm{T}_i^c 
   \to \frac{1}{2} \left( \bm{T}_i^a\bm{T}_i^b\bm{T}_i^c - \bm{T}_i^c\bm{T}_i^a\bm{T}_i^b \right)
   = \frac{i}{2} \left( f^{bcd}\,\bm{T}_i^a\bm{T}_i^d + f^{acd}\,\bm{T}_i^d\bm{T}_i^b \right) .
\end{equation}
Through the replica procedure gluon $c$ gets color-connected to either $a$ or $b$, which are part of the same cluster. The final result is thus again a fully color-connected structure. 

We have automated the replica procedure and have studied a large variety of three- and four-loop diagrams, in which gluons attach in different ways to Wilson lines. We find in all cases that only the color-connected structures shown in Figure~\ref{fig:webs} contribute to the exponent $\widetilde{S}$ and hence to the soft anomalous dimension $\bm{\Gamma}_s$. A formal proof of this result has been put forward in \cite{Gardi:2013ita} based on a generalized Baker-Campbell-Hausdorff formula.

\section{Reduction to symmetrized color structures}

One can further simplify the connected webs shown in Figure~\ref{fig:webs} by symmetrizing the attachments to the Wilson lines, as we did in \cite{Becher:2009qa}. Explicitly, the corresponding symmetrized color structures are (sums over repeated color indices are implied)\footnote{Compared with \cite{Ahrens:2012qz} we have included an extra factor of $i$ in the definition of the 5-index symbol ${\cal T}_{ijklm}$.}
\begin{equation}\label{symstruc}
\begin{aligned}
   {\cal D}_{ij} &= \bm{T}_i^a \bm{T}_j^a \equiv \bm{T}_i\cdot\bm{T}_j \,, &\quad &\text{starting at one-loop order,} \\
   {\cal T}_{ijk} &= if^{abc} \left( \bm{T}_i^a \bm{T}_j^b \bm{T}_k^c \right)_+ ,
    &\quad &\text{starting at two-loop order,} \\
   {\cal T}_{ijkl} &= f^{ade} f^{bce} \left( \bm{T}_i^a \bm{T}_j^b \bm{T}_k^c \bm{T}_l^d \right)_+ , 
    &\quad &\text{starting at three-loop order,} \\
   {\cal D}_{ijkl}^R 
   &= d_R^{abcd}\,\bm{T}_i^a \bm{T}_j^b \bm{T}_k^c \bm{T}_l^d \,,
    &\quad &\text{starting at four-loop order,} \\
   {\cal T}_{ijklm} 
   &= i f^{adf}f^{bcg}f^{efg} \left( \bm{T}_i^a \bm{T}_j^b \bm{T}_k^c \bm{T}_l^d \bm{T}_m^e \right)_+ ,
    &\quad &\text{starting at four-loop order.}
\end{aligned}
\end{equation}
Here 
\begin{equation}
   d_R^{a_1\dots a_n} = \text{Tr}_R\big( \bm{T}^{a_1}\!\ldots\bm{T}^{a_n} \big)_+
   \equiv \frac{1}{n!}\,\sum_\pi\,\text{Tr}\big( \bm{T}_R^{a_{\pi(1)}}\!\ldots\bm{T}_R^{a_{\pi(n)}} \big)
\end{equation}
are symmetric invariant tensors given in terms of traces over symmetrized products of group generators in the representation $R$. The $(\dots)_+$ prescription only acts on generators attached to the same particle line, e.g.\ ${\cal T}_{ijij}=f^{ade} f^{bce}\,(\bm{T}_i^a\bm{T}_i^c)_+ (\bm{T}_j^b \bm{T}_j^d)_+$ for $i\ne j$. For the structures ${\cal D}_{ij\dots}$ there is no need to write a $(\dots)_+$ prescription, because they are totally symmetric in their color indices. Note that (at least up to four-loop order) symmetric structures with an odd number of indices do not arise. In particular, the color-symmetric three-gluon web $d_R^{abc}\,\bm{T}_i^a \bm{T}_j^b \bm{T}_k^c$ does not appear in perturbative calculations of the three-gluon vertex function up to four-loop order \cite{Davydychev:1996pb,Davydychev:1997vh,Ruijl:2017eht}. In \cite{Ruijl:2017eht}, an argument based on Bose symmetry and charge-conjugation invariance was given that this should hold to all orders in perturbation theory. 

While the color structures ${\cal D}_{ij}$ and ${\cal D}_{ijkl}^R$ are totally symmetric in their indices, the various ${\cal T}$ structures have more complicated symmetry properties. ${\cal T}_{ijk}$ is totally antisymmetric in its indices, and it vanishes if two or three indices coincide. The structure ${\cal T}_{ijkl}$ obeys the same symmetry relations as the conformal cross ratios $\beta_{ijkl}$ in (\ref{betasym}), i.e.\
\begin{equation}
   {\cal T}_{ijkl} = {\cal T}_{jilk} = - {\cal T}_{ikjl} = - {\cal T}_{ljki} = {\cal T}_{klij} \,.
\end{equation}
It vanishes if three or four indices coincide. For two identical indices, the non-vanishing symbols are \cite{Becher:2009qa}
\begin{equation}
\begin{aligned}
   {\cal T}_{iijj} &= - {\cal T}_{ijij}
    = f^{ade} f^{bce} \left( \bm{T}_i^a \bm{T}_i^b \right)_+\! \left( \bm{T}_j^c \bm{T}_j^d \right)_+ , \\
   {\cal T}_{iijk} &= - {\cal T}_{ijik} 
    = - {\cal T}_{jiki} = {\cal T}_{jkii}
    = f^{ade} f^{bce} \left( \bm{T}_i^a \bm{T}_i^b \right)_+ \bm{T}_j^c \bm{T}_k^d \,.
\end{aligned}
\end{equation}
Useful identities for the 5-index symbol ${\cal T}_{ijklm}$ have been derived in \cite{Ahrens:2012qz}. In particular, it satisfies the relations
\begin{equation}\label{Tau5}
   {\cal T}_{ijklm} = - {\cal T}_{ikjlm} = - {\cal T}_{ljkim} = - {\cal T}_{jilkm} \,,
\end{equation}
which allow us to move any one of the first four indices to first place. Note that the fifth index is special. The ${\cal T}_{ijklm}$ symbols vanish unless at least three indices are different from each other. For the case of three different indices $i,j,k$, the symmetry properties allow us to reduce all possible structures to ${\cal T}_{iijki}$ and ${\cal T}_{iikjj}$, where the first one is antisymmetric in $j,k$, while the second one is antisymmetric in $i,j$. For the case of four different indices $i,j,k,l$, the symmetry properties imply that all structures can be reduced to ${\cal T}_{iijkl}$ and ${\cal T}_{ijkli}$, both of which are antisymmetric in $j,k$. 

Very useful additional relations can be derived using the Jacobi identity
\begin{equation}\label{Jacobi}
   f^{abe} f^{cde} + f^{ace} f^{dbe} + f^{ade} f^{bce} = 0 \,.
\end{equation}
We find
\begin{equation}\label{Jacobirels}
\begin{aligned}
   {\cal T}_{ijkl} &= {\cal T}_{ijlk} - {\cal T}_{iklj} \,, \\
   {\cal T}_{ijklm} &= {\cal T}_{ijmlk} - {\cal T}_{ikmlj} 
    = {\cal T}_{ijkml} - {\cal T}_{ljkmi} \,.
\end{aligned}
\end{equation}
The latter set of identities allows us to move the last index of the symbol ${\cal T}_{ijklm}$.

The reduction of the color factors associated with the connected webs to the symmetrized structures in (\ref{symstruc}) uses the Lie algebra $[\bm{T}^a,\bm{T}^b]=if^{abc}\,\bm{T}^c$ and the group-theory identities (recall that in the adjoint representation of the gauge group $(\bm{T}^a)_{bc}=-if^{abc}$)
\begin{equation}
\begin{aligned}
   \mbox{Tr}_A\!\left( \bm{T}^a \bm{T}^b \right)
   &= f^{acd} f^{bcd} = C_A\,\delta^{ab} \,, \\
   \mbox{Tr}_A\!\left( \bm{T}^a \bm{T}^b \bm{T}^c \right)
   &= i f^{ade} f^{beg} f^{cgd} = \frac{i C_A}{2}\,f^{abc} \,, \\[-1mm]
   \mbox{Tr}_A\!\left( \bm{T}^a \bm{T}^b \bm{T}^c \bm{T}^d \right)
   &= f^{aef} f^{bfg} f^{cgh} f^{dhe} 
    = d_A^{abcd} + \frac{C_A}{6} \left( f^{ade} f^{bce} - f^{abe} f^{cde} \right) .
\end{aligned}
\end{equation}
In deriving these expressions one uses the Jacobi identity (\ref{Jacobi}). Let us first consider the primary structure 
\begin{equation}\label{T3def}
   T_{ijk} = if^{abc}\,\bm{T}_i^a \bm{T}_j^b \bm{T}_k^c
\end{equation}
for the three-gluon web shown by the second graph in Figure~\ref{fig:webs}, where no symmetrization is applied. If all three indices $i,j,k$ are different, we obviously have $T_{ijk}={\cal T}_{ijk}$. For two different indices, we find 
\begin{equation}
   T_{iji} = - T_{iij} = - T_{ijj} = \frac{C_A}{2}\,{\cal D}_{ij} \,.
\end{equation}
If all indices are the same, then 
\begin{equation}
   T_{iii} = - \frac{C_A}{2}\,C_{R_i}\,\bm{1} \,,
\end{equation}
where $R_i$ is the color representation of the $i^{\,\rm th}$ particle. $C_F=(N_c^2-1)/(2N_c)$ and $C_A=N_c$ are the quadratic Casimir invariants in the fundamental and the adjoint representation, respectively. Hence, the color structure $T_{ijk}$ associated with the three-gluon web can be reduced to the symmetrized structure ${\cal T}_{ijk}$ and the lower-order symmetrized structures ${\cal D}_{ij}$ and $\bm{1}$.

The primary structure for the four-gluon web shown by the third graph in Figure~\ref{fig:webs} reads
\begin{equation}\label{Tdef}
   T_{ijkl} = f^{ade} f^{bce}\,\bm{T}_i^a \bm{T}_j^b \bm{T}_k^c \bm{T}_l^d \,,
\end{equation}
where again no symmetrization is applied. If all four indices $i,j,k,l$ are different, then $T_{ijkl}={\cal T}_{ijkl}$. For three different indices, we find
\begin{equation}
   T_{iijk} = - T_{ijik} = - T_{jiki} = T_{jkii} = {\cal T}_{iijk} - \frac{C_A}{4}\,{\cal T}_{ijk} \,, \qquad
   T_{ijki} = T_{jiik} = \frac{C_A}{2}\,{\cal T}_{ijk} \,.
\end{equation}
For two different indices, we obtain the relations
\begin{equation}
\begin{aligned}
   T_{iijj} &= - T_{ijij} = {\cal T}_{iijj} + \frac{C_A^2}{8}\,{\cal D}_{ij} \,, &\quad
    T_{ijji} &= - \frac{C_A^2}{4}\,{\cal D}_{ij} \,, \\
   T_{iiij} &= T_{jiii} = \frac{C_A^2}{4}\,{\cal D}_{ij} \,, &\quad 
    T_{iiji} &= - T_{ijii} = 0 \,.
\end{aligned}
\end{equation}
Finally, if all indices are the same, then
\begin{equation}
   T_{iiii} = \frac{C_A^2}{4}\,C_{R_i}\,\bm{1} \,.
\end{equation}
In all cases the primary color structure $T_{ijkl}$ can be reduced to the symmetrized structures ${\cal T}_{ijkl}$ and the lower-order symmetrized structures ${\cal T}_{ijk}$, ${\cal D}_{ij}$ and $\bm{1}$. 

As a more complicated case, we now study the four-gluon webs induced by loops of internal particles (in color representation $R$), as shown by the first two graphs in the second line of Figure~\ref{fig:webs}. They give rise to the primary color structures
\begin{equation}
   D_{ijkl}^R 
   = \mbox{Tr}_R\big(\bm{T}^a\bm{T}^b\bm{T}^c\bm{T}^d\big)\,
    \bm{T}_i^a \bm{T}_j^b \bm{T}_k^c \bm{T}_l^d \,,
\end{equation}
where the trace is taken over color generators in the representation $R$ of the gauge group. Using group-theoretic identities, one can show that \cite{vanRitbergen:1998pn}
\begin{equation}\label{eq37}
   \mbox{Tr}_R\big(\bm{T}^a\bm{T}^b\bm{T}^c\bm{T}^d\big)
   = d_R^{abcd} + \frac{i}{2}\,\big( d_R^{ade} f^{bce} - d_R^{bce} f^{ade} \big)
    + \frac{I_2(R)}{6} \left( f^{ade} f^{bce} - f^{abe} f^{cde} \right) ,
\end{equation}
where $I_2(R)$ is the second index of the representation $R$, with $I_2(F)=T_F=\frac12$ and $I_2(A)=C_A=N_c$. Note that (\ref{eq37}) introduces the 3-index symbol $d_R^{abc}$, which is known {\em not\/} to contribute to the three-gluon vertex function. However, charge-conjugation invariance ensures that, when one sums over all relevant Feynman diagrams, one always encounters the combination
\begin{equation}
   D_{ijkl}^{R,\rm sym} 
   = \frac12\,\mbox{Tr}_R\big(\bm{T}^a\bm{T}^b\bm{T}^c\bm{T}^d+\bm{T}^d\bm{T}^c\bm{T}^b\bm{T}^a\big)\,
    \bm{T}_i^a \bm{T}_j^b \bm{T}_k^c \bm{T}_l^d \,,
\end{equation}
with
\begin{equation}
   \frac12\,\mbox{Tr}_R\big(\bm{T}^a\bm{T}^b\bm{T}^c\bm{T}^d+\bm{T}^d\bm{T}^c\bm{T}^b\bm{T}^a\big)
   = d_R^{abcd} + \frac{I_2(R)}{6} \left( f^{ade} f^{bce} - f^{abe} f^{cde} \right) .
\end{equation}
It follows from this relation that
\begin{equation}\label{Dsymrela}
   D_{ijkl}^{R,\rm sym} 
   = {\cal D}_{ijkl}^R + \frac{I_2(R)}{6} \left( T_{ijkl} + T_{ilkj}  \right) 
    + C_A\,\frac{I_2(R)}{12}\,\Big[ \left( \delta_{jl} - 2\delta_{kl} \right) T_{ijk} + \delta_{jk}\,T_{ilj} \Big] \,,
\end{equation}
where the structures $T_{ijk}$ and $T_{ijkl}$ have been defined in (\ref{T3def}) and (\ref{Tdef}). Note that the 3-index $d_R^{abc}$ symbol has disappeared. Consequently, it is indeed sufficient to study the symmetrized color structures ${\cal D}_{ijkl}^R$, since the extra terms in (\ref{Dsymrela}), which have already been considered above, give rise to symmetric structures of lower order.

We finally focus on the five-gluon web shown by the last graph in Figure~\ref{fig:webs}, which gives rise to the primary color structure
\begin{equation}
   T_{ijklm} = i f^{adf}f^{bcg}f^{efg}\,\bm{T}_i^a \bm{T}_j^b \bm{T}_k^c \bm{T}_l^d \bm{T}_m^e \,.
\end{equation}
Once again, it is straightforward to show that it suffices to consider the symmetrized color structures ${\cal T}_{ijklm}$, since all commutator terms can be reduced structures already encountered in lower orders, including ${\cal T}_{ijkl}$ and ${\cal D}_{ijkl}^A$. For the purpose of illustration, we quote the relevant relations for the cases where exactly two indices coincide. We find
\begin{equation}
\begin{aligned}
   T_{ijkim} = - T_{jiikm}
   &= \frac{C_A}{2}\,{\cal T}_{ijkm} \,, \\
   T_{iiklm} = - T_{ikilm} = T_{kilim} = - T_{kliim}
   &= {\cal T}_{iiklm} - \frac{1}{2}\,{\cal D}_{iklm}^A 
     - \frac{C_A}{12} \left( {\cal T}_{ikml} + {\cal T}_{ilmk} \right) , \\
   T_{ijkli} = - T_{ljkii}
   &= {\cal T}_{ijkli} + \frac{C_A}{4}\,{\cal T}_{ijkl} \,, \\
   T_{ijklj} = - T_{ikjlj}
   &= {\cal T}_{ijklj} - \frac{C_A}{4}\,{\cal T}_{ijkl} \,. 
\end{aligned}
\end{equation}
If three or more indices coincide, the corresponding relations also contain the color structures ${\cal T}_{ijk}$, ${\cal D}_{ij}$ and $\bm{1}$. As a corollary, note that while at four-loop order in QCD the color structure ${\cal D}_{ijkl}^F$ only arises from the four-gluon vertex with an internal quark loop, the corresponding structure ${\cal D}_{ijkl}^A$ in the adjoint representation receives contributions also from diagrams without closed gluon (or ghost) loops.

\section{Anomalous dimension up to four-loop order}
\label{sec:4}

Combining the constraints imposed by soft-collinear factorization and non-abelian exponentiation, we find that the most general form of the anomalous-dimension matrix up to four-loop order can be written as
\begin{equation}\label{magic}
\begin{aligned}
   \bm{\Gamma}(\{\underline{s}\},\mu) 
   &= \sum_{(i,j)}\,\frac{\bm{T}_i\cdot\bm{T}_j}{2}\,\gamma_{\rm cusp}(\alpha_s)\,\ln\frac{\mu^2}{-s_{ij}} 
    + \sum_i\,\gamma^i(\alpha_s)\,\bm{1} \\
   &\quad\mbox{}+ f(\alpha_s) \sum_{(i,j,k)}\,{\cal T}_{iijk} 
    + \!\sum_{(i,j,k,l)}\!{\cal T}_{ijkl}\,F(\beta_{ijlk},\beta_{iklj};\alpha_s) \\[-1.5mm] 
   &\quad\mbox{}+ \sum_R g^R(\alpha_s) \bigg[ \sum_{(i,j)}\,\big( {\cal D}_{iijj}^R + 2 {\cal D}_{iiij}^R \big) 
    \ln\frac{\mu^2}{-s_{ij}}
    + \sum_{(i,j,k)} {\cal D}_{ijkk}^R\,\ln\frac{\mu^2}{-s_{ij}} \bigg] \\
   &\quad\mbox{}+ \sum_R \sum_{(i,j,k,l)}\!{\cal D}_{ijkl}^R\,G^R(\beta_{ijlk},\beta_{iklj};\alpha_s)
    + \sum_{(i,j,k,l)}\!{\cal T}_{ijkli}\,H_1(\beta_{ijlk},\beta_{iklj};\alpha_s) \\
   &\quad\mbox{}+ \sum_{(i,j,k,l,m)}\!{\cal T}_{ijklm}\,
    H_2(\beta_{ijkl},\beta_{ijmk},\beta_{ikmj},\beta_{jiml},\beta_{jlmi};\alpha_s) 
    + {\cal O}(\alpha_s^5) \,.
\end{aligned}
\end{equation}
Here $(i,j,\dots)$ refer to unordered tuples of distinct particles indices (all running from 1 to $n$). For terms involving the symmetric color structures ${\cal D}_{ijkl}^R$ we have included a sum over the color representation $R$ of the particles in the theory ($R=F,A$ for QCD).   

The first line of (\ref{magic}) contains the so-called dipole form of the anomalous dimension \cite{Becher:2009cu,Gardi:2009qi,Becher:2009qa}. The coefficients $\gamma_{\rm cusp}$ and $\gamma^i$ start at one-loop order. Note the important fact that the 3-index symbol ${\cal T}_{ijk}$ does not appear in the anomalous dimension. It would have to be multiplied by a totally antisymmetric kinematic function built out of the invariants $\beta_{ij}$, $\beta_{jk}$ and $\beta_{ki}$. The constraint (\ref{ourconstraint}) implies that this function must be linear in all three invariants. However, it is easy to show that such a function does not exist \cite{Becher:2009qa}. As a consequence, the dipole form still holds at two-loop order. 

The terms in the second line start at three-loop order and have been given in eq.~(6.17) of \cite{Becher:2009qa}. Note that the function $F(\beta_{ijlk},\beta_{iklj})$ remains invariant under the index permutations $\{ijkl\}\to\{jilk\}$ and $\{klij\}$, under which ${\cal T}_{ijkl}$ is also invariant, while $F(\beta_{ijlk},\beta_{iklj})\to F(\beta_{iklj},\beta_{ijlk})$ under the permutations $\{ijkl\}\to\{ikjl\}$ and $\{ljki\}$, under which ${\cal T}_{ijkl}$ changes sign. It follows that without loss of generality we can choose 
\begin{equation}\label{Fasym}
   F(x_1,x_2;\alpha_s) = - F(x_2,x_1;\alpha_s)
\end{equation}
to be an odd function under exchange of its arguments. This also follows more directly from the first relation in (\ref{Jacobirels}). The terms shown in the last three lines of (\ref{magic}) start at four-loop order and have been adapted from eq.~(3.16) in \cite{Ahrens:2012qz}. The antisymmetry of the color structure ${\cal T}_{ijkli}$ in $j,k$ implies that without loss of generality we can choose
\begin{equation}\label{H1symrela}
   H_1(x_1,x_2;\alpha_s) = - H_1(x_2,x_1;\alpha_s) \,.
\end{equation}
Likewise, the function $G^R$ must satisfy
\begin{equation}\label{GRsym}
   G^R(x_1,x_2;\alpha_s) = G^R(x_2,x_1;\alpha_s) = G^R(x_1-x_2,-x_2) = G^R(x_2-x_1,-x_1) \,.
\end{equation}

We emphasize the important fact that starting at four-loop order new terms involving the so-called ``cusp logarithms'' $\ln[\mu^2/(-s_{ij})]$ appear in (\ref{magic}), which are not governed by the universal cusp anomalous dimension $\gamma_{\rm cusp}(\alpha_s)$ in the first line. These terms involve new two- and three-particle color correlations. The constraint (\ref{ourconstraint}) imposed by soft-collinear factorization enforces that they appear in a certain linear combination multiplying the functions $g^R(\alpha_s)$. 

Concerning the structure of the five-particle correlations in the last line of (\ref{magic}), we note that for five different indices $i,j,k,l,m$ there exist five subsets of four indices, and in each subset one can define two linearly independent conformal cross ratios. Among these ten cross ratios there exist five linear relations \cite{Ahrens:2012qz}, which allow us to write $H_2$ as a function of five kinematic variables. With the choice made in (\ref{magic}), the relations for the other five cross ratios read
\begin{equation}
\begin{aligned}
   \beta_{iklj} &= \beta_{ikmj} + \beta_{jlmi} \,, \\
   \beta_{jklm} &= - \beta_{ijmk} + \beta_{jiml} - \beta_{jlmi} \,, \\
   \beta_{jlmk} &= - \beta_{ijkl} + \beta_{ijmk} - \beta_{ikmj} \,, \\
   \beta_{iklm} &= - \beta_{ijmk} + \beta_{ikmj} + \beta_{jiml} \,, \\
   \beta_{imkl} &= \beta_{ijkl} - \beta_{jiml} + \beta_{jlmi} \,.
\end{aligned}
\end{equation}
The symmetry relations (\ref{Tau5}) imply that, without loss of generality, we can impose the conditions
\begin{equation}\label{H2symrela}
\begin{aligned}
   H_2(y_1,y_2,y_3,y_4,y_5;\alpha_s) &= - H_2(y_1,y_4,y_5,y_2,y_3;\alpha_s) \\
   &= - H_2(-y_1,y_3,y_2,y_4-y_2+y_3,y_5+y_1-y_2+y_3;\alpha_s) \\
   &= - H_2(-y_1,y_2-y_4+y_5,y_3+y_1-y_4+y_5,y_5,y_4;\alpha_s) \,,
\end{aligned}
\end{equation}
the first one of which is particularly simple. 

In the color-space formalism, color conservation translates into the statement that one gets zero when summing over the particle index of the right-most color generator in a given color structure, i.e.\
\begin{equation}\label{colorcons}
   \dots\,\sum_{i=1}^n\,\bm{T}_i^a = 0 \,.
\end{equation}
We can use this relation to derive some additional non-trivial conditions on the function $H_2$, which are based on the color identities (for five different indices $i,j,k,l,m$)
\begin{equation}\label{nicesums}
\begin{aligned}
   \sum_{m\ne i,j,k,l}\,{\cal T}_{ijklm}
   &= - {\cal T}_{ijkli} - {\cal T}_{ijklj} - {\cal T}_{ijklk} - {\cal T}_{ijkll} \,, \\
   \sum_{l\ne i,j,k,m}\,{\cal T}_{ijklm}
   &= - {\cal T}_{ijkjm} - {\cal T}_{ijkkm} - {\cal T}_{ijkmm} \,.
\end{aligned}
\end{equation}
To derive these identities, we have expressed ${\cal T}_{ijklm}$ in terms of products of color generators contracted with $f^{abc}$ symbols, performed the sums over $m$ and $l$ using (\ref{colorcons}) after moving the corresponding color generators all the way to the right, and rewritten the answer in terms of symmetrized color structures. It is remarkable that, contrary to (\ref{faltern}) below, no lower-order color structures appear in these relations, even though they appear in intermediate steps of the calculation. Analogous sums over the indices $i,j,k$ can be derived from the second relation through the symmetry relations (\ref{Tau5}), which allow us to move any one of the first four indices to fourth place. Consider now the following color sums:
\begin{equation}
   S_1 = \!\sum_{(i,j,k,l,m)}\!\!{\cal T}_{ijklm}\,H(\beta_{ijlk},\beta_{iklj}) \,, \qquad
   S_2 = \!\sum_{(i,j,k,l,m)}\!\!{\cal T}_{ijklm}\,H(\beta_{ijmk},\beta_{ikmj}) 
\end{equation}
with some function $H(x_1,x_2)$. The antisymmetry of the color symbols under $j\leftrightarrow k$ implies that we can impose the condition $H(x_1,x_2)=-H(x_2,x_1)$. The two sums are defined such that one summation index does not appear in the arguments of the function $H$, so this index can be summed over using color conservation. Using the corresponding expressions in (\ref{nicesums}) and renaming some summation indices, we find that
\begin{equation}\label{Ssums}
   S_1 = 0 \,, \qquad
   S_2 = \sum_{(i,j,k,l)}\!\Big[ - {\cal T}_{ijkli}\,H(\beta_{ijlk},\beta_{iklj}) 
    + {\cal T}_{iijkl}\,\hat H(\beta_{ijlk},\beta_{iklj}) \Big] \,,
\end{equation}
where the new function $\hat H$ is related to $H$ by
\begin{equation}\label{hatHdef}
   \hat H(x_1,x_2) = H(x_2-x_1,-x_1) - H(x_1-x_2,-x_2) = - \hat H(x_2,x_1) \,.
\end{equation}
To proceed further, we use the second relation in (\ref{Jacobirels}), which follows from the Jacobi identity (\ref{Jacobi}). Setting $m=i$, this relation gives ${\cal T}_{ijkli}={\cal T}_{iiklj}-{\cal T}_{iijlk}$, which can be used to derive that
\begin{equation}\label{5Tsimpl}
   \sum_{(i,j,k,l)}\!{\cal T}_{ijkli}\,H(\beta_{ijlk},\beta_{iklj}) 
   = \sum_{(i,j,k,l)}\!{\cal T}_{iijkl}\,\hat H(\beta_{ijlk},\beta_{iklj}) \,.
\end{equation}
Combining this result with the second relation in (\ref{Ssums}), we find that 
\begin{equation}
   S_2 = 0 \,. 
\end{equation}
The fact that $S_1=S_2=0$ shows that any contribution to the function $H_2$ which only depends on a subset of four different particle indices gives a vanishing result and can be dropped. In this sense, the function $H_2$ parametrizes genuine five-particle correlation terms.

In our master formula (\ref{magic}) we have cleaned up the notation compared with the original expressions given in our earlier papers \cite{Becher:2009qa,Ahrens:2012qz} and we have used some color identities to eliminate two structures arising at four-loop order. The precise relations between the coefficient functions in (\ref{magic}) and those used in our previous work can be found in Appendix~\ref{app:A}. Note that expression (\ref{magic}) for the anomalous dimension can be rewritten in equivalent ways using color conservation. For example, the term proportional to $f$ in the second line could be recast into the form
\begin{equation}\label{faltern}
   f(\alpha_s) \sum_{(i,j,k)}\,{\cal T}_{iijk}
   = - f(\alpha_s) \sum_{(i,j)}\,{\cal T}_{iijj} + \frac{C_A^2}{8}\,f(\alpha_s) \sum_i C_{R_i}\,\bm{1} \,,
\end{equation}
where the latter term can be absorbed into the one-particle anomalous dimensions $\gamma^i$ in (\ref{magic}). We prefer to keep the original form on the left-hand side of (\ref{faltern}), because it shows that $f$ only contributes if there are at least three different particles involved in the process. Likewise, using (\ref{5Tsimpl}) with $H=H_1$, the term proportional to the function $H_1$ in (\ref{magic}) could be rewritten in the alternative form 
\begin{equation}\label{eqHrel}
   \sum_{(i,j,k,l)}\!{\cal T}_{ijkli}\,H_1(\beta_{ijlk},\beta_{iklj})
   = \sum_{(i,j,k,l)}\!{\cal T}_{iijkl}\,\hat H_1(\beta_{ijlk},\beta_{iklj}) \,.
\end{equation}

We close this section with an important remark. Using conformal transformations, a relation between soft functions for multi-parton scattering at small transverse momentum and soft functions arising in jet processes was established in \cite{Vladimirov:2016dll}. Based on this relation, it was argued that only irreducible color structures containing an even number of color generators $\bm{T}_i$ can appear in the soft anomalous-dimension matrix $\bm{\Gamma}_s$ \cite{Vladimirov:2017ksc}. While the absence of color structures with three generators is a simple consequence of the symmetry properties of the associated coefficient function (see above) \cite{Becher:2009cu,Gardi:2009qi,Becher:2009qa,Dixon:2009ur}, the observation made in  \cite{Vladimirov:2017ksc} -- if true -- would imply that the functions $H_1$ and $H_2$ in (\ref{magic}) vanish identically. The relation derived in \cite{Vladimirov:2016dll} applies to soft functions which can be written as matrix elements of time-ordered products of Wilson lines. Such soft functions are associated with inclusive cross sections and much more restricted than the general amplitude-level soft functions considered here. It is therefore not obvious to us that the relation in \cite{Vladimirov:2016dll} is sufficient to exclude the five-particle correlations in \eqref{magic}.

\section{Coefficient functions and cusp anomalous dimensions}
\label{sec:ff}

Thanks to the efforts of several groups, much is known about the various coefficient functions entering the anomalous dimension in (\ref{magic}). Remarkably, the universal cusp anomalous dimension $\gamma_{\rm cusp}(\alpha_s)$ is known at four-loop order. The expansion coefficients up to three-loop order are given in Appendix~\ref{app:B}. Except for two constants, which are presently only available in numerical form, the four-loop coefficient $\gamma_3^{\rm cusp}$ is known analytically. From the calculations of the cusp anomalous dimension in the large-$N_c$ limit performed in \cite{Henn:2016men,Lee:2016ixa}, combined with the calculation of the $n_f C_F^2$ terms in \cite{Grozin:2018vdn} and the evaluation of the contributions involving quartic Casimir invariants in \cite{Moch:2017uml,Moch:2018wjh,Lee:2019zop,Henn:2019rmi}, one can determine the terms proportional to $C_A^3$ as well as $n_f C_F C_A$ and $n_f C_A^2$. The contributions proportional to $n_f^2 C_F$ and $n_f^2 C_A$ were obtained in \cite{Davies:2016jie,Lee:2017mip,vonManteuffel:2019wbj}, while those proportional to $n_f^3$ were first calculated in \cite{Beneke:1995pq}. Combining all these ingredients, we find
\begin{equation}\label{gcusp3}
\begin{aligned}
   \gamma_3^{\rm cusp}
   &= C_A^3\!\left( \frac{84278}{81} - \frac{44416\pi^2}{243} + \frac{20992\zeta_3}{27} + \frac{902\pi^4}{45} 
    - 352\zeta _5 - \frac{292\pi^6}{315} - \frac{176\pi^2\zeta_3}{9} - 32\zeta_3^2 - \frac{k_1}{12} \right) \\
   &\quad\mbox{}+ 2T_F n_f\,\bigg[ C_F^2 \left( \frac{572}{9} + \frac{592\zeta_3}{3} - 320\zeta_5 \right)
    + C_F C_A\,k_2 \\
   &\qquad\mbox{}+ C_A^2 \left( - \frac{41170}{81} + \frac{13130\pi^2}{243} - \frac{17536\zeta_3}{27} 
    - \frac{44\pi^4}{27} + \frac{2816\zeta_5}{9} + \frac{128\pi^2\zeta_3}{9} - \frac{k_2}{2} \right) \bigg] \\
   &\quad\mbox{}+ (2T_F n_f)^2 \left[ C_F \left( \frac{2392}{81} - \frac{640\zeta_3}{9} + \frac{16\pi^4}{45} \right)
    + C_A \left( \frac{923}{81} - \frac{304\pi^2}{243} + \frac{2240\zeta_3}{27} - \frac{56\pi^4}{135} \right) \right] \\
   &\quad\mbox{}+ (2T_F n_f)^3 \left( - \frac{32}{81} + \frac{64\zeta_3}{27} \right) \\
   &\approx (610.26\pm 0.1)\,C_A^3 - 31.0554\,n_f C_F^2 + (38.75\pm 0.2)\,n_f C_F C_A - (440.64\pm 0.1)\,n_f C_A^2 \\
   &\quad\mbox{}- 21.3144\,C_F n_f^2 + 58.3674\,n_f^2 C_A + 2.45426\,n_f^3 \,, 
\end{aligned}
\end{equation}
where the constants 
\begin{equation}\label{kivals}
   k_1 = - (253.5\pm 1.0) \,, \qquad
   k_2 = 38.75\pm 0.2
\end{equation}
have been obtained numerically in \cite{Moch:2017uml,Moch:2018wjh}. Note that $k_1$ is related to quartic Casimir invariants, which do not contribute to the universal cusp anomalous dimension $\gamma^{\rm cusp}$ but to the cusp anomalous dimensions of quarks and gluons, see (\ref{casimir4}) and (\ref{Mochres}) below. Recently, the conjecture
\begin{equation}
   k_2 = - \frac{34066}{81} + \frac{220\pi^2}{9} + \frac{3712\zeta_3}{9} - \frac{88\pi^4}{45}
    + 160\zeta_5 - \frac{64\pi^2\zeta_3}{3} 
   \approx 38.7954
\end{equation}
was presented in \cite{Bruser:2019auj}, which is in excellent agreement with the numerical result in (\ref{kivals}). When this is used, the coefficient of the $n_f C_A^2$ term in (\ref{gcusp3}) becomes
\begin{equation}
   - \frac{24137}{81} + \frac{10160\pi^2}{243} - \frac{23104\zeta_3}{27} 
    - \frac{88\pi^4}{135} + \frac{2096\zeta_5}{9} + \frac{224\pi^2\zeta_3}{9} 
   \approx - 440.667 \,.
\end{equation}

The single-particle anomalous dimensions $\gamma^i$ for quarks and gluons ($i=q,g$) are known to three-loop order and are given in Appendix~\ref{app:B}. Explicit expressions for the function $F(x_1,x_2;\alpha_s)$ and the coefficient $f(\alpha_s)$ can be derived from the three-loop results for the soft anomalous dimension for three-particle amplitudes obtained in the pioneering paper \cite{Almelid:2015jia}. This yields
\begin{equation}\label{eq33}
\begin{aligned}
   F(x_1,x_2;\alpha_s) &= 2\,{\cal F}(e^{x_1},e^{x_2}) \left( \frac{\alpha_s}{4\pi} \right)^3 
    + {\cal O}(\alpha_s^4) \,, \\
   f(\alpha_s) &= 16 \left( \zeta_5 + 2\zeta_2\zeta_3 \right) \left( \frac{\alpha_s}{4\pi} \right)^3 
    + {\cal O}(\alpha_s^4) \,,
\end{aligned}
\end{equation}
where the function ${\cal F}(x,y)$ can be expressed in terms of Brown's single-valued harmonic polylogarithms \cite{Brown:2004ugm,Dixon:2012yy}. Defining a complex variable $z$ such that $z\bar z=x$ and $(1-z)(1-\bar z)=y$, one finds that ${\cal F}(x,y)={\cal L}(1-z)-{\cal L}(z)$, where
\begin{equation}
   {\cal L}(z) = {\cal L}_{10101}(z) + 2\zeta_2 \left[ {\cal L}_{001}(z) + {\cal L}_{100}(z) \right] .
\end{equation}

Of the remaining terms in (\ref{magic}), which start at four-loop order, only the coefficients $g^R$ can be determined from presently available calculations. To this end, we exploit the fact that the anomalous dimension $\bm{\Gamma}$ simplifies drastically for the case of $n=2$ particles. We obtain (with $i=q,g$)
\begin{equation}\label{Gff}
   \bm{\Gamma}(s_{12},\mu) 
   = - \bigg[ C_{R_i} \gamma_{\rm cusp}(\alpha_s) + 2 \sum_R g^R(\alpha_s)\,{\cal D}_{iiii}^R \bigg] 
    \ln\frac{\mu^2}{-s_{12}} + 2\gamma^i(\alpha_s) + {\cal O}(\alpha_s^5) \,,
\end{equation}
where the right-hand side is proportional to the unit matrix in color space, and from here on we omit the symbol $\bm{1}$ to indicate such terms. For $i=q,g$ these quantities are the anomalous dimensions of the quark and gluon form factors. The structure
\begin{equation}
   {\cal D}_{iiii}^R = d_R^{abcd}\,\bm{T}_i^a \bm{T}_i^b \bm{T}_i^c \bm{T}_i^d
   = d_R^{abcd} \left( \bm{T}^a \bm{T}^b \bm{T}^c \bm{T}^d \right)_{R_i}
   \equiv C_4(R_i,R)
\end{equation}
defines a quartic Casimir invariant, which commutes with all generators in the representation $R$ of the gauge group. If $R$ is irreducible, then Schur's lemma implies that $C_4(R_i,R)$ is proportional to the unit matrix. One finds
\begin{equation}
   C_4(R_i,R) = \frac{d_{R_i}^{abcd} d_R^{abcd}}{N_{R_i}} 
   \equiv \frac{d_{R_i R}^{(4)}}{N_{R_i}} \,,
\end{equation}
where the symbol $d_{R_i R}^{(4)}$ was introduced in \cite{Moch:2018wjh}, and $N_{R_i}$ is the dimension of the representation $R_i$ (with $N_F=N_c$ and $N_A=N_c^2-1$). For an $SU(N_c)$ gauge theory the relevant combinations are (we use $T_F=\frac12$)
\begin{equation}\label{dsymbols}
\begin{aligned}
   d_{FF}^{(4)} &= \frac{(N_c^4-6N_c^2+18)(N_c^2-1)}{96 N_c^2} \,, \\
   d_{FA}^{(4)} &= d_{AF}^{(4)} = \frac{N_c(N_c^2+6)(N_c^2-1)}{48} \,, \\
   d_{AA}^{(4)} &= \frac{N_c^2(N_c^2+36)(N_c^2-1)}{24} \,.
\end{aligned}
\end{equation}

The coefficient of the logarithm in (\ref{Gff}) is called the cusp anomalous dimension for particle $i$, which should be distinguished from the universal cusp anomalous dimension $\gamma_{\rm cusp}$. We find
\begin{equation}\label{casimir4}
   \Gamma_{\rm cusp}^i(\alpha_s) 
   = C_{R_i} \gamma_{\rm cusp}(\alpha_s) + 2\sum_R C_4(R_i,R)\,g^R(\alpha_s) + {\cal O}(\alpha_s^5) \,.
\end{equation}
The four-loop terms proportional to the quartic Casimir invariants $C_4(R_i,R)$ violate the simple (quadratic) Casimir scaling relation $\Gamma_{\rm cusp}^q(\alpha_s)/C_F=\Gamma_{\rm cusp}^g(\alpha_s)/C_A$. Indeed, using arguments based on the AdS/CFT correspondence, results for the cusp anomalous dimension obtained in the strong-coupling limit were known to be inconsistent with simple Casimir scaling for a long time \cite{Armoni:2006ux,Alday:2007hr,Alday:2007mf}. It was also found recently that simple Casimir scaling is violated in ${\cal N}=4$ supersymmetric Yang-Mills theory \cite{Boels:2017skl,Boels:2017ftb}. Here we have shown that these terms still obey a generalized form of Casimir scaling, meaning that the same two functions $g^F$ and $g^A$ appear in both $\Gamma_{\rm cusp}^q$ and $\Gamma_{\rm cusp}^g$, and their weights are governed by the quartic Casimir invariants $C_4(R_i,R)$. This fact has first been observed in \cite{Moch:2018wjh}, where the authors have obtained the four-loop coefficients of the coefficients $g^R$ for QCD in numerical form. The coefficient $g^F$ has later also been calculated analytically \cite{Lee:2019zop,Henn:2019rmi}. Using these results, we find
\begin{equation}\label{Mochres}
\begin{aligned}
   g^F(\alpha_s) &= T_F n_f\,\bigg( \frac{128\pi^2}{3} - \frac{256\zeta_3}{3}
    - \frac{1280\zeta_5}{3} \bigg) \left( \frac{\alpha_s}{4\pi} \right)^4 + {\cal O}(\alpha_s^5) \,, \\
   g^A(\alpha_s) &= (-253.5\pm 1.0) \left( \frac{\alpha_s}{4\pi} \right)^4 + {\cal O}(\alpha_s^5) \,.
\end{aligned}
\end{equation}
The numerical coefficient in the second result coincides with the constant $k_1$ in (\ref{kivals}). 

Note that in the large-$N_c$ limit the ratio
\begin{equation}
   \frac{C_4(A,R)}{C_4(F,R)} = 2 + {\cal O}\bigg(\frac{1}{N_c^2}\bigg)
\end{equation} 
becomes independent of the representation $R$, and it approaches the same limiting value as the ratio $C_A/C_F$ \cite{Dixon:2017nat}. As a result, in this limit the quark and gluon cusp anomalous dimensions {\em do\/} obey the simple Casimir scaling relation at least up to four-loop order,
\begin{equation}
   \lim_{N_c\to\infty}\,\frac{\Gamma_{\rm cusp}^g(\alpha_s)}{\Gamma_{\rm cusp}^q(\alpha_s)} \Bigg|_{\rm 4-loop}\!
   = \lim_{N_c\to\infty}\,\frac{C_A}{C_F} = 2 \,.
\end{equation} 

As a final remark, let us mention that, using arguments based on conformal symmetry, collinear factorization and the Regge limit, the authors of \cite{Almelid:2017qju} were able to ``bootstrap'' the three-loop expression for the function $F(x_1,x_2;\alpha_s)$ in (\ref{eq33}) up to an overall rational factor. It would be interesting to explore whether similar arguments allow one to determine (or constrain) the unknown four-loop functions $G^R$, $H_1$ and $H_2$ in (\ref{magic}).

\section{Two-particle collinear limits}
\label{sec:5}

The result (\ref{magic}) can be constrained further by studying two-particle collinear limits. The conformal cross ratios $\beta_{ijkl}$ either vanish or diverge when two of the four particle momenta become collinear. In order to study the collinear limit properly, we consider the case in which the momenta of particles 1 and 2 are almost aligned with each other, such that \cite{Becher:2009qa}
\begin{equation}
   p_1^\mu = zE n^\mu + p_\perp^\mu 
    - \frac{p_\perp^2}{4zE}\,\bar n^\mu \,, \qquad
   p_2^\mu = (1-z)E n^\mu - p_\perp^\mu 
    - \frac{p_\perp^2}{4(1-z)E}\,\bar n^\mu \,, 
\end{equation}
where $n^2=\bar n^2=0$ and $n\cdot\bar n=2$, and the ratio $p_\perp/E$ serves as a small expansion parameter. This parameterization is such that $p_1^2=p_2^2=0$ remain on-shell, while $-s_{12}=p_\perp^2/[z(1-z)]$. The collinear limit corresponds to taking $p_\perp\to 0$ at fixed energy $E$. 

After a lengthy calculation, we find that in the limit where particles 1 and 2 (both assumed to be outgoing) become collinear our result (\ref{magic}) implies the following contribution to the anomalous dimension of the splitting amplitude in (\ref{collconstraint}): 
\begin{equation}\label{splitresu}
\begin{aligned}
   \bm{\Gamma}_{\rm Sp}(\{p_1,p_2\},\mu)
   &= \gamma_{\rm cusp}(\alpha_s)\,\bigg\{ \bm{T}_1\cdot\bm{T}_2 
    \left[ \ln\frac{\mu^2}{-s_{12}} + \ln z(1-z) \right] 
    + C_{R_1} \ln z + C_{R_2} \ln(1-z) \bigg\} \\[1mm]
   &\quad\mbox{}+ \gamma^1(\alpha_s) + \gamma^2(\alpha_s) - \gamma^P(\alpha_s) \\[1mm]
   &\quad\mbox{}- f(\alpha_s)\,\bigg[ 2 {\cal T}_{1122} - 4\,\sum_{i\ne 1,2}\,{\cal T}_{12ii} 
    + \frac{C_A^2}{4}\,\bm{T}_1\cdot\bm{T}_2 \bigg] 
    + \!\sum_{(i,j)\ne 1,2}\!8 {\cal T}_{12ij}\,F(\omega_{ij},0;\alpha_s) \\[-1mm]
   &\quad\mbox{}+ \sum_R\,g^R(\alpha_s)\,\bigg\{ \Big[ 6{\cal D}_{1122}^R + 4 \big( {\cal D}_{1112}^R 
    + {\cal D}_{1222}^R \big) \Big] \left[ \ln\frac{\mu^2}{-s_{12}} + \ln z(1-z) \right] \\
   &\hspace{3.1cm}\mbox{}+ 2 \big[ {\cal D}_{1111}^R \ln z + {\cal D}_{2222}^R \ln(1-z) \big]
    + \!\sum_{(i,j)\ne 1,2}\!2{\cal D}_{12ij}^R\,\omega_{ij} \bigg\} \\[-1mm]
   &\quad\mbox{}+ \sum_R \sum_{(i,j)\ne 1,2}\!12 {\cal D}_{12ij}^R\,G^R(\omega_{ij},0;\alpha_s) \\
   &\quad\mbox{}+ \text{contributions involving ${\cal T}_{ijklm}$ symbols} + {\cal O}(\alpha_s^5) \,,
\end{aligned}
\end{equation}
where $\gamma^P$ is the anomalous dimension associated with the unresolved particle $P$. We have defined the quantity (at leading non-trivial order in $p_\perp/E$)
\begin{equation}\label{eq72}
   \omega_{ij} \equiv \beta_{12ij}
   = \ln\frac{p_\perp^2}{4z^2(1-z)^2E^2} + \ln\frac{(-s_{ij})}{(-n\cdot p_i)(-n\cdot p_j)}\to - \infty
\end{equation}
and used that $\epsilon_{ij}\equiv\beta_{1ij2}={\cal O}(p_\perp/E)$ vanishes in the collinear limit. We have also used the symmetry properties (\ref{Fasym}) and (\ref{GRsym}). The appearance of color generators for particles other than 1 and 2 in the anomalous dimension of the splitting amplitude would violate collinear factorization, and hence the corresponding structures must vanish in the above result. 

Let us focus first on the terms in the third line. It was assumed in \cite{Becher:2009qa} and \cite{Dixon:2009ur} that the coefficients of the terms violating collinear factorization vanish individually, i.e.\ $f(\alpha_s)=0$ and $F(\omega_{ij},0;\alpha_s)\to 0$ for $\omega_{ij}\to-\infty$. There is, however, a more general solution based on the color identity
\begin{equation}
   \sum_{(i,j)\ne 1,2}\!{\cal T}_{12ij}
   = - \sum_{i\ne 1,2} {\cal T}_{12ii} - {\cal T}_{1122} - \frac{C_A^2}{8}\,\bm{T}_1\cdot\bm{T}_2 \,. 
\end{equation}
If we impose the condition
\begin{equation}
   \lim_{\omega\to-\infty} F(\omega,0;\alpha_s) = \frac{f(\alpha_s)}{2} \,, 
\end{equation}
then collinear factorization holds. The explicit expression for $F$ obtained in \cite{Almelid:2015jia} shows that this condition is indeed satisfied. 

Concerning the terms shown in the next three lines, we had assumed in \cite{Becher:2009qa,Ahrens:2012qz} that the coefficients of the terms involving particle indices other than 1 and 2 vanish individually, i.e.\ $g^R(\alpha_s)=0$ and $G^R(\omega_{ij},0;\alpha_s)\to 0$ for $\omega_{ij}\to-\infty$. Under this assumption, the cusp anomalous dimension in (\ref{casimir4}) would obey Casimir scaling at four-loop order. Once again, there exists a more general solution, in which we impose that the function $G^R$ of conformal cross ratios obeys the relation 
\begin{equation}\label{gRlimit}
   \lim_{\omega\to-\infty} G^R(\omega,0;\alpha_s) 
   = - \frac{g^R(\alpha_s)}{6}\,\omega \,,
\end{equation}
meaning that it diverges logarithmically in the collinear limit. The coefficients $g^R(\alpha_s)$ are then no longer forced to vanish, in accordance with the explicit results in (\ref{Mochres}). 

Let us finally comment on the terms in (\ref{splitresu}) involving the 5-index ${\cal T}_{ijklm}$ symbols, whose explicit form is discussed in Appendix~\ref{app:C}. There are various contributions to the anomalous dimension of the splitting amplitude descending from the functions $H_1$ and $H_2$, see (\ref{5Tsplittingres2}). The requirement that the sum of these terms must not depend on particle indices other than 1 and 2 implies the condition 
\begin{equation}\label{h1limit}
   \lim_{\omega\to-\infty}\,H_1(\omega,0;\alpha_s) = 0
\end{equation}
as well as a more non-trivial relation given in (\ref{nontrivial}). We find that when these relations are satisfied, the contributions involving the 5-index ${\cal T}_{ijklm}$ symbols vanish identically.

Combining all pieces, we conclude that up to four-loop order the anomalous dimension of the splitting amplitude is given by
\begin{equation}\label{funrela}
\begin{aligned}
   &\bm{\Gamma}_{\rm Sp}(\{p_1,p_2\},\mu) \\
   &= \bigg\{ \gamma_{\rm cusp}(\alpha_s)\,\bm{T}_1\cdot\bm{T}_2 
    + \sum_R 2 g^R(\alpha_s)\,\Big[ 3{\cal D}_{1122}^R + 2 \big( {\cal D}_{1112}^R + {\cal D}_{1222}^R \big) \Big]
    \bigg\} \left[ \ln\frac{\mu^2}{-s_{12}} + \ln z(1-z) \right] \\
   &\quad\mbox{}+ \gamma_{\rm cusp}(\alpha_s)\,\Big[ C_{R_1} \ln z + C_{R_2} \ln(1-z) \Big]
    + \gamma^1(\alpha_s) + \gamma^2(\alpha_s) - \gamma^P(\alpha_s) \\
   &\quad\mbox{}- 6 f(\alpha_s) \left( {\cal T}_{1122} + \frac{C_A^2}{8}\,\bm{T}_1\cdot\bm{T}_2 \right)    
    + \sum_R 2 g^R(\alpha_s)\,\Big[ {\cal D}_{1111}^R \ln z + {\cal D}_{2222}^R \ln(1-z) \Big] 
    + {\cal O}(\alpha_s^5) \,.
\end{aligned}
\end{equation}
This result holds irrespectively of whether or not the five-particle contributions proportional to $H_1$ and $H_2$  contribute to the anomalous dimension \eqref{magic} (see the discussion at the end of Section \ref{sec:4}).

\section{Applications}
\label{sec:6}

The most important accomplishment of our analysis is that it provides explicit and complete expressions for the anomalous-dimension matrices needed to perform resummations of large logarithms in $n$-jet cross sections with next-to-next-to-next-to-leading logarithmic (N$^3$LL) accuracy. At this order, one resums logarithms of the form $\alpha_s^n L^k$ with $(n-2)\le k\le 2n$ in the logarithm of a cross section. This requires that one knows the logarithmically enhanced terms in the anomalous dimension (the so-called ``cusp logarithms'') to four-loop order and the remaining terms to three-loop accuracy. The appearance of cusp logarithms is a characteristic feature of anomalous dimensions associated with amplitudes sensitive to Sudakov double logarithms. Note that N$^3$LL resummation is what is needed to perform a consistent matching onto NNLO fixed-order expressions for the cross sections, which is becoming state-of-the-art in perturbative QCD. From our general result (\ref{magic}), we obtain
\begin{equation}
\begin{aligned}
   \bm{\Gamma}(\{\underline{s}\},\mu) 
   &= \sum_{(i,j)}\,\frac{\bm{T}_i\cdot\bm{T}_j}{2}\,\gamma_{\rm cusp}(\alpha_s)\,\ln\frac{\mu^2}{-s_{ij}} \\[-1.5mm]
   &\quad\mbox{}+ \sum_R g^R(\alpha_s) \bigg[ \sum_{(i,j)}\,\big( {\cal D}_{iijj}^R + 2 {\cal D}_{iiij}^R \big) 
    \ln\frac{\mu^2}{-s_{ij}} + \sum_{(i,j,k)} {\cal D}_{ijkk}^R\,\ln\frac{\mu^2}{-s_{ij}} \bigg] \\
   &\quad\mbox{}+ \sum_i\,\gamma^i(\alpha_s) + f(\alpha_s) \sum_{(i,j,k)}\,{\cal T}_{iijk} 
    + \!\sum_{(i,j,k,l)}\!{\cal T}_{ijkl}\,F(\beta_{ijlk},\beta_{iklj};\alpha_s) \\[-1.5mm]
   &\quad\mbox{}+ {\cal O}\bigg(\alpha_s^4,\alpha_s^5\ln\frac{\mu^2}{-s_{ij}}\bigg) .
\end{aligned}
\end{equation}
Based on our analysis, the terms involving cusp logarithms are now known to four-loop order, while the remaining contributions in the third line are known to three-loop order.

As a second application, we briefly consider the important case of processes involving only a small number of external particles. While the form-factor case ($n=2$) has already been discussed in Section~\ref{sec:ff}, we now study the case of three particles ($n=3$). This is relevant for resumming large QCD corrections to important collider processes such as $e^+ e^-\to\text{3 jets}$ (which involves $e^+ e^-\to q\bar q g$ at the parton level) and $pp\to H+\text{jet}$ (which involves $q\bar q\to Hg$, $qg\to Hq$ and $gg\to Hg$ at the parton level). For the special case of three-particle amplitudes, many of the multi-particle correlations do not contribute, and other terms can be simplified using color conservation. We find that the general form of the anomalous dimension in (\ref{magic}) reduces to
\begin{equation}
\begin{aligned}
   \bm{\Gamma}(\{\underline{s}\},\mu) 
   &= \frac{\gamma_{\rm cusp}(\alpha_s)}{2} \left[ \left( C_{R_3} - C_{R_1} - C_{R_2} \right) 
    \ln\frac{\mu^2}{(-s_{12})} + \text{cyclic permutations} \right] \\[-0.5mm]
   &\quad\mbox{}+ \gamma^1(\alpha_s) + \gamma^2(\alpha_s) + \gamma^3(\alpha_s) 
    + \frac{C_A^2}{8}\,f(\alpha_s) \left( C_{R_1} + C_{R_2} +C_{R_3} \right) \\[0.5mm]
   &\quad\mbox{}+ \sum_{(i,j)}\,\bigg[ - f(\alpha_s)\,{\cal T}_{iijj} 
    + \sum_R g^R(\alpha_s)\,\big( 3 {\cal D}_{iijj}^R + 4 {\cal D}_{iiij}^R \big) \ln\frac{\mu^2}{-s_{ij}} \bigg]
    + {\cal O}(\alpha_s^5) \,,
\end{aligned}
\end{equation}
where $C_{R_i}$ are the quadratic Casimir invariants of the three particles. Starting at three-loop order non-trivial color structures appear, which cannot be simplified further.

\section{Conclusions}

Using techniques based on soft-collinear factorization in SCET and the non-abelian exponentiation theorem for matrix elements of soft Wilson-line correlators, we have derived the general form of the anomalous dimension $\bm{\Gamma}$ governing the IR divergences of $n$-particle scattering amplitudes in massless, non-abelian gauge theories up to four-loop order. Our result for $\bm{\Gamma}$ has been given in (\ref{magic}). Exploiting non-trivial color identities, we have significantly simplified the general form compared with previous proposals in the literature by eliminating two structures in the four-loop result. We find that the four-loop contribution involves three new color structures multiplying cusp logarithms, which describe color correlations among two or three particles and whose form is completely determined by a single constant coefficient $g^R(\alpha_s)$ for each representation $R$ of the gauge group. For QCD, these coefficients can be determined from existing calculations of the IR divergences of the quark and gluon form factors. In addition, three yet unknown functions of conformal cross ratios account for four-particle ($G^R$ and $H_1$) and five-particle ($H_2$) correlations in color and kinematics. 

The fact that in the limit where two particles become collinear the anomalous dimension must obey the relation (\ref{collconstraint}) puts highly non-trivial constraints on the functional form of the coefficient functions, which depend on the external particle's momenta through so-called conformal cross ratios. By carefully reevaluating these constraints, we find that at four-loop order color structures involving contractions of totally symmetric $d_R^{abcd}$ tensors appear along with cusp logarithms $\ln[\mu^2/(-s_{ij})]$. As a consequence, naive Casimir scaling of the cusp anomalous dimensions associated with the quark and gluon form factors is violated, while a generalized form of Casimir scaling still holds.

It has recently been shown that the three-loop expression for the function $F$ in (\ref{eq33}) can be derived, up to an overall rational factor, using arguments based on conformal symmetry, collinear factorization and the Regge limit \cite{Almelid:2017qju}. It may be possible to derive in an analogous way expressions for the functions $G^R$ and $H_1$, which like $F$ depend on a pair of conformal cross ratios with the same four indices. 

Our results provide for a better understanding of the intricate pattern of IR divergences of scattering amplitudes in non-abelian gauge theories. At the same time, they are also important from a practical point of view. The anomalous dimension we have derived provides an important ingredient for the resummation of large (Sudakov) logarithms in $n$-jet processes at N$^3$LL accuracy. At this order, one needs the cusp logarithms in the anomalous dimension to four-loop order and the remaining terms at three-loop level. All of these ingredients are provided by our analysis independently of the number of external particles. While the functions $G^R$, $H_1$ and $H_2$ describing multi-particle correlations at four-loop order and higher are not yet known, our results provide the complete four-loop anomalous dimensions for amplitudes with up to three color-charged particles, once the quark and gluon form factors are known to this order. This will provide non-trivial consistency checks on amplitude computations for such important processes as $e^+ e^-\to\text{3 jets}$ and $pp\to H+\text{jet}$. 

\subsection*{Acknowlegdements}

We thank Eric Laenen, Sven Moch, Maximilian Stahlhofen, Alexey Vladimirov and Andreas Vogt for useful discussions. We are grateful to the Mainz Institute for Theoretical Physics (MITP) of the Cluster of Excellence PRISMA$^+$\! (project ID 39083149), funded by the German Research Foundation (DFG), for hospitality and support during the final stages of this research. This work of M.N.\ is also supported by grant 05H18UMCA1 of the German Federal Ministry for Education and Research (BMBF). The research of T.B.\ is supported by the Swiss National Science Foundation (SNF) under grant  200020\_182038. 

\subsection*{Note added}
While this paper was under review, the preprint \cite{Henn:2019swt} appeared, in which  a fully analytical result for the cusp anomalous dimension was presented. This fixes the constant
\begin{equation*}
k_1 = -\frac{32 \pi ^2}{3} +\frac{64 \zeta _3}{3} +\frac{1760 \zeta _5}{3} - \frac{496 \pi^6}{945} -192 \zeta _3^2 \approx -253.332 \,
   \end{equation*}
given in numerical form in \eqref{kivals} in the main text.

\begin{appendix}
\renewcommand{\theequation}{A.\arabic{equation}}
\setcounter{equation}{0}

\section{Connection of (\ref{magic}) with results in the literature}
\label{app:A}

In our master formula (\ref{magic}) we have cleaned up the notation compared with the original expressions given in our earlier papers. Our functions $f$ and $F$ are related to the corresponding objects in \cite{Becher:2009qa} by
\begin{equation}
\begin{aligned}
   f(\alpha_s) &= - \bar f_2(\alpha_s) \,, \\
   F(\beta_{ijlk},\beta_{iklj};\alpha_s) 
   &= F(\beta_{ijkl},\beta_{iklj}-\beta_{iljk}) - \frac{\bar f_1(\alpha_s)}{4}\,\beta_{ijkl} \,.
\end{aligned}
\end{equation}
Furthermore, our functions $g^R$, $G^R$ and $H_i$ are connected with the corresponding quantities in \cite{Ahrens:2012qz} (where we did not write out the representation index $R$ explicitly) by
\begin{equation}
\begin{aligned}
   g^R(\alpha_s) &= - g_1(\alpha_s) \,, \\[-1.5mm]
   G^R(\beta_{ijlk},\beta_{iklj};\alpha_s) 
   &= G_1(\beta_{ijkl},\beta_{iklj},\beta_{iljk}) + \frac{g_5(\alpha_s)}{3} \,, \\
   H_1(\beta_{ijlk},\beta_{iklj};\alpha_s) 
   &= -i G_2(\beta_{ijkl},\beta_{iklj}) + 2i G_3(\beta_{ijlk},\beta_{ilkj}) \,, \\[2mm]
   H_2(\beta_{ijkl},\beta_{ijmk},\beta_{ikmj},\beta_{jiml},\beta_{jlmi};\alpha_s) 
   &= -i G_4(\beta_{ijkl},\beta_{iklj},\beta_{ijkm},\beta_{ikmj},\beta_{ijml}) \,.
\end{aligned}
\end{equation}
Contrary to the original definitions of the functions $G_2$, $G_3$ and $G_4$, the new functions $H_1$ and $H_2$ are defined such that their imaginary parts correspond to physical discontinuities. 

A non-trivial aspect of the above relations involves the identity 
\begin{equation}
   \sum_{(i,j)}\,{\cal D}_{iijj}^R 
   = \frac13 \sum_{(i,j,k,l)} {\cal D}_{ijkl}^R + \sum_i\,{\cal D}_{iiii}^R \,,
\end{equation}
which allows us to absorb the contribution involving $g_5$ in eq.~(3.16) of \cite{Ahrens:2012qz} into the function $G^R$. The extra terms proportional to ${\cal D}_{iiii}^R$ can be absorbed into the one-particle anomalous dimensions $\gamma^i$ in (\ref{magic}). Another non-trivial relation has been given in (\ref{5Tsimpl}), which allows us to relate the contributions proportional to the functions $G_2$ or $G_3$ in eq.~(3.16) of \cite{Ahrens:2012qz} to each other and absorb them into a single function $H_1$.

\renewcommand{\theequation}{B.\arabic{equation}}
\setcounter{equation}{0}

\section{Anomalous-dimension coefficients and $\bm{Z}$-factor}
\label{app:B}

Given a UV renormalized, on-shell $n$-particle scattering amplitude $|{\cal M}_n(\epsilon,\{\underline s\})\rangle$ with IR divergences regularized in $d=4-2\epsilon$ dimensions, one can obtain the finite amplitude $|{\cal M}_n(\{\underline s\},\mu)\rangle$, in which all IR are subtracted in a minimal way, from the relation \cite{Becher:2009cu}
\begin{equation}
   |{\cal M}_n(\{\underline s\},\mu)\rangle 
   = \lim_{\epsilon\to 0}\,\bm{Z}^{-1}(\epsilon,\{\underline s\},\mu)\,|{\cal M}_n(\epsilon,\{\underline s\})\rangle \,.
\end{equation}
The $\bm{Z}$ factor is related to the anomalous dimension $\bf\Gamma$ studied in the present paper by 
\begin{equation}
   {\bf\Gamma}(\{\underline s\},\mu) 
   = - \bm{Z}^{-1}(\epsilon,\{\underline s\},\mu)\,\frac{d}{d\ln\mu}\,\bm{Z}(\epsilon,\{\underline s\},\mu) \,.
\end{equation}
Up to four-loop order, the solution to this equation was derived in \cite{Becher:2009qa,Ahrens:2012qz}. One obtains
\begin{equation}
\begin{aligned}
   \ln{\bf Z} &= \frac{\alpha_s}{4\pi}
    \left( \frac{\Gamma_0'}{4\epsilon^2} + \frac{{\bf\Gamma}_0}{2\epsilon} \right)
    + \left( \frac{\alpha_s}{4\pi} \right)^2 \left( - \frac{3\beta_0 \Gamma_0'}{16\epsilon^3}
    + \frac{\Gamma_1'-4\beta_0 {\bf\Gamma}_0}{16\epsilon^2}
    + \frac{{\bf\Gamma}_1}{4\epsilon} \right) \\
   &\quad\mbox{}+ \left( \frac{\alpha_s}{4\pi} \right)^3
    \bigg( \frac{11\beta_0^2\Gamma_0'}{72\epsilon^4}
    - \frac{5\beta_0\Gamma_1'+8\beta_1\Gamma_0'-12\beta_0^2{\bf\Gamma}_0}{72\epsilon^3} 
    + \frac{\Gamma_2'-6\beta_0{\bf\Gamma}_1-6\beta_1{\bf\Gamma}_0}{36\epsilon^2}
    + \frac{{\bf\Gamma}_2}{6\epsilon} \bigg) \\
   &\quad\mbox{}+ \left( \frac{\alpha_s}{4\pi} \right)^4
    \bigg( - \frac{25\beta_0^3\Gamma_0'}{192\epsilon^5}
    + \frac{13\beta_0^2\Gamma_1'+40\beta_0\beta_1\Gamma_0'-24\beta_0^3{\bf\Gamma}_0}{192\epsilon^4} \\ 
   &\hspace{25mm}\mbox{}- \frac{7\beta_0\Gamma_2'+9\beta_1\Gamma_1'+15\beta_2\Gamma_0'
    -24\beta_0^2{\bf\Gamma}_1-48\beta_0\beta_1{\bf\Gamma}_0}{192\epsilon^3} \\
   &\hspace{25mm}\mbox{}+ 
    \frac{\Gamma_3'-8\beta_0{\bf\Gamma}_2-8\beta_1{\bf\Gamma}_1-8\beta_2{\bf\Gamma}_0}{64\epsilon^2}
    + \frac{{\bf\Gamma}_3}{8\epsilon} \bigg) + {\cal O}(\alpha_s^5) \,, 
\end{aligned}
\end{equation}
where we have expanded the anomalous dimension and $\beta$-function as 
\begin{equation}\label{Gbexp}
   {\bf\Gamma}(\alpha_s) = \sum_{n=0}^{\infty}\,{\bf\Gamma}_n \left( \frac{\alpha_s}{4\pi} \right)^{n+1} ,
    \qquad
   \beta(\alpha_s) = -2\alpha_s \sum_{n=0}^{\infty}\,\beta_n \left( \frac{\alpha_s}{4\pi} \right)^{n+1} ,
\end{equation}
and similarly for the function
\begin{equation}
   \Gamma'(\alpha_s) = \frac{\partial}{\partial\ln\mu}\,\bm{\Gamma}(\{\underline s\},\mu)
   = - \sum_i \Gamma_{\rm cusp}^i(\alpha_s) \,, 
\end{equation}
where the cusp anomalous dimensions $\Gamma_{\rm cusp}^i(\alpha_s)$ have been given in (\ref{casimir4}). Through relations (\ref{magic}) and (\ref{casimir4}), the coefficients $\bm{\Gamma}_n$ and $\Gamma_n'$ can in turn be expressed in terms of the expansion coefficients of the anomalous dimensions $\gamma_{\rm cusp}$, $\gamma^q$ and $\gamma^g$, as well as of the coefficient functions of the higher-order terms, all defined in analogy with the first relation in (\ref{Gbexp}). 

We now list the expansion coefficients of the quantities $\gamma_{\rm cusp}$, $\gamma^q$ and $\gamma^g$ up to three-loop order in the $\overline{{\rm MS}}$ renormalization scheme. The coefficients of the universal cusp anomalous dimension $\gamma_{\rm cusp}$ are given by \cite{Moch:2004pa}
\begin{equation}
\begin{aligned}
   \gamma_0^{\rm cusp} &= 4 \,, \\
   \gamma_1^{\rm cusp} &= \left( \frac{268}{9} - \frac{4\pi^2}{3} \right) C_A - \frac{80}{9}\,T_F n_f \,, \\
   \gamma_2^{\rm cusp} &= C_A^2 \left( \frac{490}{3} - \frac{536\pi^2}{27}
    + \frac{44\pi^4}{45} + \frac{88}{3}\,\zeta_3 \right) 
    + C_A T_F n_f  \left( - \frac{1672}{27} + \frac{160\pi^2}{27} - \frac{224}{3}\,\zeta_3 \right) \\
   &\quad\mbox{}+ C_F T_F n_f \left( - \frac{220}{3} + 64\zeta_3 \right) - \frac{64}{27}\,T_F^2 n_f^2 \,.
\end{aligned}
\end{equation}
The anomalous dimension $\gamma^q=\gamma^{\bar q}$ can be determined from the three-loop expression for the divergent part of the on-shell quark form factor in QCD \cite{Moch:2005id}. One obtains \cite{Becher:2006mr}
\begin{equation}
\begin{aligned}
   \gamma_0^q &= -3 C_F \,, \\
   \gamma_1^q &= C_F^2 \left( -\frac{3}{2} + 2\pi^2 - 24\zeta_3 \right)
    + C_F C_A \left( - \frac{961}{54} - \frac{11\pi^2}{6} + 26\zeta_3 \right)
    + C_F T_F n_f \left( \frac{130}{27} + \frac{2\pi^2}{3} \right) , \\
   \gamma_2^q &= C_F^3 \left( -\frac{29}{2} - 3\pi^2 - \frac{8\pi^4}{5} - 68\zeta_3
    + \frac{16\pi^2}{3}\,\zeta_3 + 240\zeta_5 \right) \\
   &\quad\mbox{}+ C_F^2 C_A \left( - \frac{151}{4} + \frac{205\pi^2}{9} + \frac{247\pi^4}{135}
    - \frac{844}{3}\,\zeta_3 - \frac{8\pi^2}{3}\,\zeta_3 - 120\zeta_5 \right) \\
   &\quad\mbox{}+ C_F C_A^2 \left( - \frac{139345}{2916} - \frac{7163\pi^2}{486} - \frac{83\pi^4}{90} 
    + \frac{3526}{9}\,\zeta_3 - \frac{44\pi^2}{9}\,\zeta_3 - 136\zeta_5 \right) \\
   &\quad\mbox{}+ C_F^2 T_F n_f \left( \frac{2953}{27} - \frac{26\pi^2}{9} - \frac{28\pi^4}{27}
    + \frac{512}{9}\,\zeta_3 \right) \\
   &\quad\mbox{}+ C_F C_A T_F n_f \left( - \frac{17318}{729} + \frac{2594\pi^2}{243} + \frac{22\pi^4}{45} 
    - \frac{1928}{27}\,\zeta_3 \right) \\
   &\quad\mbox{}+ C_F T_F^2 n_f^2 \left( \frac{9668}{729} - \frac{40\pi^2}{27} - \frac{32}{27}\,\zeta_3 \right) .
\end{aligned}
\end{equation}
Similarly, the expression for the gluon anomalous dimension can be extracted from the divergent part of the gluon form factor obtained in \cite{Moch:2005id}. One finds \cite{Becher:2009qa}
\begin{equation}
\begin{aligned}
   \gamma_0^g &= - \beta_0 = - \frac{11}{3}\,C_A + \frac43\,T_F n_f \,, \\
   \gamma_1^g &= C_A^2 \left( -\frac{692}{27} + \frac{11\pi^2}{18} + 2\zeta_3 \right) 
    + C_A T_F n_f \left( \frac{256}{27} - \frac{2\pi^2}{9} \right) + 4 C_F T_F n_f \,, 
 \end{aligned}
\end{equation}   
 \begin{equation}
\begin{aligned}   
   \gamma_2^g &= C_A^3 \left( - \frac{97186}{729} + \frac{6109\pi^2}{486} - \frac{319\pi^4}{270} 
    + \frac{122}{3}\,\zeta_3 - \frac{20\pi^2}{9}\,\zeta_3 - 16\zeta_5 \right) 
    \hspace{8mm} \\
   &\quad\mbox{}+ C_A^2 T_F n_f \left( \frac{30715}{729} - \frac{1198\pi^2}{243} + \frac{82\pi^4}{135} 
    + \frac{712}{27}\,\zeta_3 \right) \\
   &\quad\mbox{}+ C_A C_F T_F n_f \left( \frac{2434}{27} - \frac{2\pi^2}{3} - \frac{8\pi^4}{45}
    - \frac{304}{9}\,\zeta_3 \right) - 2 C_F^2 T_F n_f \\
   &\quad\mbox{}+ C_A T_F^2 n_f^2 \left( - \frac{538}{729} + \frac{40\pi^2}{81} - \frac{224}{27}\,\zeta_3 \right) 
    - \frac{44}{9}\,C_F T_F^2 n_f^2 \,.
\end{aligned}
\end{equation}
Our results for $\gamma^q$ and $\gamma^g$ are valid in the conventional dimensional regularization scheme, where polarization vectors and spinors of all particles are treated as $d$-dimensional objects, so that gluons have $(2-2\epsilon)$ helicity states. At two-loop order, the corresponding expressions in the 't~Hooft-Veltman scheme \cite{tHooft:1972tcz}, the dimensional reduction scheme \cite{Siegel:1979wq} and the four-dimensional helicity scheme \cite{Bern:1991aq} have been calculated in \cite{Broggio:2015dga}.

\renewcommand{\theequation}{C.\arabic{equation}}
\setcounter{equation}{0}

\section[Contributions of 5-index color structures to $\bm{\Gamma}_{\rm Sp}$]{Contributions from 5-index color structures to $\bm{\Gamma}_{\rm Sp}$}
\label{app:C}

It is straightforward but tedious to work out the contributions to the anomalous dimension of the splitting amplitude originating from the terms proportional to the 5-index ${\cal T}$ symbols in (\ref{magic}). We find 
\begin{equation}\label{5Tsplitting}
\begin{aligned}
   \bm{\Gamma}_{\rm Sp}(\{p_1,p_2\},\mu)
   &\ni - \!\sum_{(j,k,l)\ne 1,2}\!{\cal T}_{1jkl2}\,H_1(\beta_{1jlk},\beta_{1klj};\alpha_s) \\
   &\quad\mbox{}+ 2\!\sum_{(k,l)\ne 1,2}\!\big( {\cal T}_{12kl1} - {\cal T}_{12klk} \big)\,
    H_1(\omega_{kl},0;\alpha_s) \\
   &\quad\mbox{}+ \sum_{(j,k)\ne 1,2}\!\big( 2 {\cal T}_{1jk21} - {\cal T}_{1jk2j} - {\cal T}_{1jk2k} \big)\,
    H_1(-\omega_{jk},-\omega_{jk};\alpha_s) \\
   &\quad\mbox{}+ 4\!\sum_{(k,l,m)\ne 1,2}\!{\cal T}_{12klm}\,
    H_2(\omega_{kl},\omega_{km},0,\omega_{lm},0;\alpha_s) \\
   &\quad\mbox{}+ 2\!\sum_{(j,k,m)\ne 1,2}\!{\cal T}_{1jk2m}\,
    H_2(0,\beta_{1jmk},\beta_{1kmj},-\omega_{jm},-\omega_{jm};\alpha_s) \\
   &\quad\mbox{}+ 4\!\sum_{(j,k,l)\ne 1,2}\!{\cal T}_{1jkl2}\,
    H_2(\beta_{1jkl},-\omega_{jk},-\omega_{jk},0,\omega_{jl};\alpha_s) \\
   &\quad\mbox{}+ (1\leftrightarrow 2) \,.
\end{aligned}
\end{equation}
In a first step, one finds a rather long expression for this result, due to the many different ways in which one can distribute the index pair $(1,2)$ onto the color structure ${\cal T}_{ijklm}$. We have simplified the answer using the symmetry properties of the functions $H_1$ and $H_2$ given in (\ref{H1symrela}) and (\ref{H2symrela}) along with the identities
\begin{equation}
   \beta_{1ijk} = \beta_{2ijk} = \omega_{jk} - \omega_{ik} \,,
\end{equation}
which hold up to terms of ${\cal O}(p_\perp/E)$. Moreover, the terms in third and fifth lines vanish owing to (\ref{H1symrela}) and the third equation in (\ref{H2symrela}). 

We can simplify the result (\ref{5Tsplitting}) further using the Jacobi identity, which implies the relations shown in the second line of (\ref{Jacobirels}). This allows us to rewrite ${\cal T}_{1jkl2}={\cal T}_{12klj}-{\cal T}_{12jlk}$. Given that the antisymmetry under exchange of $j\leftrightarrow k$ is already built into the symmetry properties of the functions $H_1$ and $H_2$ in the first and last lines of (\ref{5Tsplitting}), we can use instead ${\cal T}_{1jkl2}\to 2{\cal T}_{12klj}$ and group three of the terms together to obtain
\begin{equation}\label{5Tsplittingres2}
\begin{aligned}
   \bm{\Gamma}_{\rm Sp}(\{p_1,p_2\},\mu)
   &\ni \!\sum_{(k,l,m)\ne 1,2}\!\big( {\cal T}_{12klm} + {\cal T}_{21klm}\big)\,\Big[ 
    - 2 H_1(\beta_{1mlk},\beta_{1klm};\alpha_s) \\[-3mm]
   &\hspace{10mm}\mbox{}+ 4 H_2(\omega_{kl},\omega_{km},0,\omega_{lm},0;\alpha_s) 
    + 8 H_2(\beta_{1mkl},-\omega_{km},-\omega_{km},0,\omega_{lm};\alpha_s) \Big] \\    
   &\quad\mbox{}+ 2\!\sum_{(k,l)\ne 1,2}\!\Big[ \big( {\cal T}_{12kl1} - {\cal T}_{12klk} \big)
    + (1\leftrightarrow 2) \Big]\,H_1(\omega_{kl},0;\alpha_s) \,.
\end{aligned}
\end{equation}

Several non-trivial cancellations need to take place in order for the various terms in (\ref{5Tsplittingres2}) not to depend on particle indices other than 1 and 2. In particular, the first term inside the bracket in the first line, which involves a non-trivial kinematic function $H_1(\beta_{1mlk},\beta_{1klm};\alpha_s)$, needs to cancel against the two remaining terms inside the bracket. Such a cancellation is indeed possible, because $\omega_{km}=\omega_{kl}+\beta_{1lmk}$ and $\omega_{lm}=\omega_{kl}+\beta_{1kml}$. We can thus rewrite the terms inside the bracket as
\begin{equation}
\begin{aligned}
   \big[ \dots \big]
   &= - 2 H_1(\beta_{1mlk},\beta_{1klm};\alpha_s)
    + 4 H_2(\omega_{kl},\omega_{kl}+\beta_{1lmk},0,\omega_{kl}+\beta_{1kml},0;\alpha_s) \\
   &\quad\mbox{}+ 8 H_2(-\beta_{1kml},-\omega_{kl}-\beta_{1lmk},-\omega_{kl}-\beta_{1lmk},
                        0,\omega_{kl}+\beta_{1kml};\alpha_s) \,,
\end{aligned}
\end{equation}
where $\omega_{kl}\to-\infty$ while the conformal cross ratios stay fixed. The arguments of $H_1$ can be related to those of the other functions by $\beta_{1mlk}=-\beta_{1lmk}$ and $\beta_{1klm}=\beta_{1kml}-\beta_{1lmk}$. The cancellation mentioned above does not need to be complete. All we need to require is\footnote{While it would be reasonable to expect that in the collinear limit the function $K$ approaches a finite function $K_0(\beta_1,\beta_2;\alpha_s)$, we cannot exclude the possibility that it contains divergent terms proportional to powers of $\omega$, in analogy with (\ref{gRlimit}).} 
\begin{equation}\label{nontrivial}
\begin{aligned}
   \lim_{\omega_{kl}\to-\infty} \Big[
   & - H_1(-\beta_{1lmk},\beta_{1kml}-\beta_{1lmk};\alpha_s)
    + 2 H_2(\omega_{kl},\omega_{kl}+\beta_{1lmk},0,\omega_{kl}+\beta_{1kml},0;\alpha_s) \\[-1.5mm]
   &\mbox{}+ 4 H_2(-\beta_{1kml},-\omega_{kl}-\beta_{1lmk},-\omega_{kl}-\beta_{1lmk},
                   0,\omega_{kl}+\beta_{1kml};\alpha_s) \Big] \\
   &\hspace{-3.5mm}= K(\beta_{1kml},\beta_{1lmk},\omega_{kl};\alpha_s) \,,
\end{aligned}
\end{equation}
where the right-hand side must be symmetric under the exchange of $k$ and $l$. In order words, the function $K$ can be arbitrary, as long as it satisfies
\begin{equation}
   K(\beta_1,\beta_2,\omega;\alpha_s) = K(\beta_2,\beta_1,\omega;\alpha_s) \,.
\end{equation}
That this is a sufficient condition follows from the fact that 
\begin{equation}
   {\cal T}_{12klm} + {\cal T}_{21klm} = {\cal T}_{12klm} - {\cal T}_{12lkm}
\end{equation}
is antisymmetric under $k\leftrightarrow l$, and hence the first sum in (\ref{5Tsplittingres2}) evaluates to zero as long as (\ref{nontrivial}) holds. 

For the term in the last line of (\ref{5Tsplittingres2}) we must require that the function $H_1(\omega_{kl},0;\alpha_s)$ becomes independent of $\omega_{kl}$ in the collinear limit. But this is not enough, since after a lengthy calculation we find that
\begin{equation}
   \sum_{(k,l)\ne 1,2}\!\Big[ \big( {\cal T}_{12kl1} - {\cal T}_{12klk} \big) + (1\leftrightarrow 2) \Big] 
   = \sum_{k\ne 1,2} 2 \big( {\cal T}_{112kk} + {\cal T}_{221kk} \big)
\end{equation}
cannot be reduced to an expression that only depends on the particle indices 1 and 2. Hence, we must require that the stronger condition (\ref{h1limit}) holds. It then follows that the right-hand side of (\ref{5Tsplittingres2}) vanishes. Hence, the structures involving 5-index ${\cal T}$ symbols in (\ref{magic}) do not contribute to the anomalous dimension of the splitting amplitude.

\end{appendix}


\begin{thebibliography}{99}

\bibitem{Becher:2009cu} 
  T.~Becher and M.~Neubert,
  Phys.\ Rev.\ Lett.\  {\bf 102}, 162001 (2009)
  [Erratum: Phys.\ Rev.\ Lett.\  {\bf 111}, 199905 (2013)]
  [arXiv:0901.0722 [hep-ph]].
  
\bibitem{Korchemskaya:1994qp} 
  I.~A.~Korchemskaya and G.~P.~Korchemsky,
  Nucl.\ Phys.\ B {\bf 437}, 127 (1995)
  [hep-ph/9409446].

\bibitem{Bauer:2000yr} 
  C.~W.~Bauer, S.~Fleming, D.~Pirjol and I.~W.~Stewart,
  Phys.\ Rev.\ D {\bf 63}, 114020 (2001)
  [hep-ph/0011336].
  
\bibitem{Bauer:2001yt} 
  C.~W.~Bauer, D.~Pirjol and I.~W.~Stewart,
  Phys.\ Rev.\ D {\bf 65}, 054022 (2002)
  [hep-ph/0109045].
  
\bibitem{Bauer:2002nz} 
  C.~W.~Bauer, S.~Fleming, D.~Pirjol, I.~Z.~Rothstein and I.~W.~Stewart,
  Phys.\ Rev.\ D {\bf 66}, 014017 (2002)
  [hep-ph/0202088].

\bibitem{Beneke:2002ph} 
  M.~Beneke, A.~P.~Chapovsky, M.~Diehl and T.~Feldmann,
  Nucl.\ Phys.\ B {\bf 643}, 431 (2002)
  [hep-ph/0206152].

\bibitem{Catani:1996vz} 
  S.~Catani and M.~H.~Seymour,
  Nucl.\ Phys.\ B {\bf 485}, 291 (1997)
  [Erratum: Nucl.\ Phys.\ B {\bf 510}, 503 (1998)]
  [hep-ph/9605323].

\bibitem{Becher:2003kh} 
  T.~Becher, R.~J.~Hill, B.~O.~Lange and M.~Neubert,
  Phys.\ Rev.\ D {\bf 69}, 034013 (2004)
  [hep-ph/0309227].

\bibitem{Korchemskaya:1992je} 
  I.~A.~Korchemskaya and G.~P.~Korchemsky,
  Phys.\ Lett.\ B {\bf 287}, 169 (1992).

\bibitem{Gardi:2009qi} 
  E.~Gardi and L.~Magnea,
  JHEP {\bf 0903}, 079 (2009)
  [arXiv:0901.1091 [hep-ph]].
    
\bibitem{Becher:2009qa} 
  T.~Becher and M.~Neubert,
  JHEP {\bf 0906}, 081 (2009)
  [Erratum: JHEP {\bf 1311}, 024 (2013)]
  [arXiv:0903.1126 [hep-ph]].

\bibitem{Gatheral:1983cz} 
  J.~G.~M.~Gatheral,
  Phys.\ Lett.\  {\bf 133B}, 90 (1983).

\bibitem{Frenkel:1984pz} 
  J.~Frenkel and J.~C.~Taylor,
  Nucl.\ Phys.\ B {\bf 246}, 231 (1984).

\bibitem{Gardi:2010rn} 
  E.~Gardi, E.~Laenen, G.~Stavenga and C.~D.~White,
  JHEP {\bf 1011}, 155 (2010)
  [arXiv:1008.0098 [hep-ph]].

\bibitem{Gardi:2013ita} 
  E.~Gardi, J.~M.~Smillie and C.~D.~White,
  JHEP {\bf 1306}, 088 (2013)
  [arXiv:1304.7040 [hep-ph]].

\bibitem{Aybat:2006wq} 
  S.~M.~Aybat, L.~J.~Dixon and G.~F.~Sterman,
  Phys.\ Rev.\ Lett.\  {\bf 97}, 072001 (2006)
  [hep-ph/0606254].

\bibitem{Aybat:2006mz} 
  S.~M.~Aybat, L.~J.~Dixon and G.~F.~Sterman,
  Phys.\ Rev.\ D {\bf 74}, 074004 (2006)
  [hep-ph/0607309].

\bibitem{Becher:2009kw} 
  T.~Becher and M.~Neubert,
  Phys.\ Rev.\ D {\bf 79}, 125004 (2009)
  [Erratum: Phys.\ Rev.\ D {\bf 80}, 109901 (2009)]
  [arXiv:0904.1021 [hep-ph]].

\bibitem{Ferroglia:2009ep} 
  A.~Ferroglia, M.~Neubert, B.~D.~Pecjak and L.~L.~Yang,
  Phys.\ Rev.\ Lett.\  {\bf 103}, 201601 (2009)
  [arXiv:0907.4791 [hep-ph]].

\bibitem{Ferroglia:2009ii} 
  A.~Ferroglia, M.~Neubert, B.~D.~Pecjak and L.~L.~Yang,
  JHEP {\bf 0911}, 062 (2009)
  [arXiv:0908.3676 [hep-ph]].

\bibitem{Ahrens:2012qz} 
  V.~Ahrens, M.~Neubert and L.~Vernazza,
  JHEP {\bf 1209}, 138 (2012)
  [arXiv:1208.4847 [hep-ph]].
    
\bibitem{Almelid:2015jia} 
  \O.~Almelid, C.~Duhr and E.~Gardi,
  Phys.\ Rev.\ Lett.\  {\bf 117}, 172002 (2016)
  [arXiv:1507.00047 [hep-ph]].

\bibitem{Berends:1988zn} 
  F.~A.~Berends and W.~T.~Giele,
  Nucl.\ Phys.\ B {\bf 313}, 595 (1989).

\bibitem{Mangano:1990by} 
  M.~L.~Mangano and S.~J.~Parke,
  Phys.\ Rept.\  {\bf 200}, 301 (1991)
  [hep-th/0509223].

\bibitem{Bern:1995ix} 
  Z.~Bern and G.~Chalmers,
  Nucl.\ Phys.\ B {\bf 447}, 465 (1995)
  [hep-ph/9503236].

\bibitem{Kosower:1999xi} 
  D.~A.~Kosower,
  Nucl.\ Phys.\ B {\bf 552}, 319 (1999)
  [hep-ph/9901201].

\bibitem{Catani:2011st} 
  S.~Catani, D.~de Florian and G.~Rodrigo,
  JHEP {\bf 1207}, 026 (2012)
  [arXiv:1112.4405 [hep-ph]].

\bibitem{Forshaw:2012bi} 
  J.~R.~Forshaw, M.~H.~Seymour and A.~Siodmok,
  JHEP {\bf 1211}, 066 (2012)
  [arXiv:1206.6363 [hep-ph]].

\bibitem{Bret:2011xm} 
  V.~Del Duca, C.~Duhr, E.~Gardi, L.~Magnea and C.~D.~White,
  Phys.\ Rev.\ D {\bf 85}, 071104 (2012)
  [arXiv:1108.5947 [hep-ph]].

\bibitem{DelDuca:2011ae} 
  V.~Del Duca, C.~Duhr, E.~Gardi, L.~Magnea and C.~D.~White,
  JHEP {\bf 1112}, 021 (2011)
  [arXiv:1109.3581 [hep-ph]].
    
\bibitem{Mitov:2010rp} 
  A.~Mitov, G.~Sterman and I.~Sung,
  Phys.\ Rev.\ D {\bf 82}, 096010 (2010)
  [arXiv:1008.0099 [hep-ph]].

\bibitem{Gardi:2011wa} 
  E.~Gardi and C.~D.~White,
  JHEP {\bf 1103}, 079 (2011)
  [arXiv:1102.0756 [hep-ph]].

\bibitem{Gardi:2011yz} 
  E.~Gardi, J.~M.~Smillie and C.~D.~White,
  JHEP {\bf 1109}, 114 (2011)
  [arXiv:1108.1357 [hep-ph]].

\bibitem{Vladimirov:2014wga} 
  A.~A.~Vladimirov,
  Phys.\ Rev.\ D {\bf 90}, 066007 (2014)
  [arXiv:1406.6253 [hep-th]].
  
\bibitem{Vladimirov:2015fea} 
  A.~A.~Vladimirov,
  JHEP {\bf 1506}, 120 (2015)
  [arXiv:1501.03316 [hep-th]].

\bibitem{Laenen:2008gt} 
  E.~Laenen, G.~Stavenga and C.~D.~White,
  JHEP {\bf 0903}, 054 (2009)
  [arXiv:0811.2067 [hep-ph]].

\bibitem{spinGlass}
  M. Mezard, G. Parisi, and M. Virasoro, {\it Spin Glass Theory and Beyond: An Introduction to the Replica Method and 
  Its Applications}, World Scientific (1987) 476\,pp.
 
\bibitem{Davydychev:1996pb} 
  A.~I.~Davydychev, P.~Osland and O.~V.~Tarasov,
  Phys.\ Rev.\ D {\bf 54}, 4087 (1996)
  [Erratum: Phys.\ Rev.\ D {\bf 59}, 109901 (1999)]
  [hep-ph/9605348].
  
\bibitem{Davydychev:1997vh} 
  A.~I.~Davydychev, P.~Osland and O.~V.~Tarasov,
  Phys.\ Rev.\ D {\bf 58}, 036007 (1998)
  [hep-ph/9801380].

\bibitem{Ruijl:2017eht} 
  B.~Ruijl, T.~Ueda, J.~A.~M.~Vermaseren and A.~Vogt,
  JHEP {\bf 1706}, 040 (2017)
  [arXiv:1703.08532 [hep-ph]].

\bibitem{vanRitbergen:1998pn} 
  T.~van Ritbergen, A.~N.~Schellekens and J.~A.~M.~Vermaseren,
  Int.\ J.\ Mod.\ Phys.\ A {\bf 14}, 41 (1999)
  [hep-ph/9802376].

\bibitem{Vladimirov:2016dll} 
  A.~A.~Vladimirov,
  Phys.\ Rev.\ Lett.\  {\bf 118}, 062001 (2017)
  [arXiv:1610.05791 [hep-ph]].

\bibitem{Vladimirov:2017ksc} 
  A.~Vladimirov,
  JHEP {\bf 1804}, 045 (2018)
  [arXiv:1707.07606 [hep-ph]].
  
\bibitem{Dixon:2009ur} 
  L.~J.~Dixon, E.~Gardi and L.~Magnea,
  JHEP {\bf 1002}, 081 (2010)
  [arXiv:0910.3653 [hep-ph]].


\bibitem{Henn:2016men} 
  J.~M.~Henn, A.~V.~Smirnov, V.~A.~Smirnov and M.~Steinhauser,
  JHEP {\bf 1605}, 066 (2016)
  [arXiv:1604.03126 [hep-ph]].

\bibitem{Lee:2016ixa} 
  J.~Henn, A.~V.~Smirnov, V.~A.~Smirnov, M.~Steinhauser and R.~N.~Lee,
  JHEP {\bf 1703}, 139 (2017)
  [arXiv:1612.04389 [hep-ph]].

\bibitem{Grozin:2018vdn} 
  A.~Grozin,
  JHEP {\bf 1806}, 073 (2018)  
  [Addendum: JHEP {\bf 1901}, 134 (2019)]
  [arXiv:1805.05050 [hep-ph]].

\bibitem{Moch:2017uml} 
  S.~Moch, B.~Ruijl, T.~Ueda, J.~A.~M.~Vermaseren and A.~Vogt,
  JHEP {\bf 1710}, 041 (2017)
  [arXiv:1707.08315 [hep-ph]].

\bibitem{Moch:2018wjh} 
  S.~Moch, B.~Ruijl, T.~Ueda, J.~A.~M.~Vermaseren and A.~Vogt,
  Phys.\ Lett.\ B {\bf 782}, 627 (2018)
  [arXiv:1805.09638 [hep-ph]].

\bibitem{Lee:2019zop} 
  R.~N.~Lee, A.~V.~Smirnov, V.~A.~Smirnov and M.~Steinhauser,
  JHEP {\bf 1902}, 172 (2019)
  [arXiv:1901.02898 [hep-ph]].

\bibitem{Henn:2019rmi} 
  J.~M.~Henn, T.~Peraro, M.~Stahlhofen and P.~Wasser,
  Phys.\ Rev.\ Lett.\  {\bf 122}, 201602 (2019)
  [arXiv:1901.03693 [hep-ph]].

\bibitem{Davies:2016jie} 
  J.~Davies, A.~Vogt, B.~Ruijl, T.~Ueda and J.~A.~M.~Vermaseren,
  Nucl.\ Phys.\ B {\bf 915}, 335 (2017)
  [arXiv:1610.07477 [hep-ph]].

\bibitem{Lee:2017mip} 
  R.~N.~Lee, A.~V.~Smirnov, V.~A.~Smirnov and M.~Steinhauser,
  Phys.\ Rev.\ D {\bf 96}, 014008 (2017)
  [arXiv:1705.06862 [hep-ph]].

\bibitem{vonManteuffel:2019wbj} 
  A.~von Manteuffel and R.~M.~Schabinger,
  Phys.\ Rev.\ D {\bf 99}, 094014 (2019)
  [arXiv:1902.08208 [hep-ph]].

\bibitem{Beneke:1995pq} 
  M.~Beneke and V.~M.~Braun,
  Nucl.\ Phys.\ B {\bf 454}, 253 (1995)
  [hep-ph/9506452].
  
\bibitem{Bruser:2019auj} 
  R.~Br\"user, A.~Grozin, J.~M.~Henn and M.~Stahlhofen,
  JHEP {\bf 1905}, 186 (2019)
  [arXiv:1902.05076 [hep-ph]].
  
\bibitem{Brown:2004ugm} 
  F.~C.~S.~Brown,
  Compt.\ Rend.\ Math.\  {\bf 338}, 527 (2004).

\bibitem{Dixon:2012yy} 
  L.~J.~Dixon, C.~Duhr and J.~Pennington,
  JHEP {\bf 1210}, 074 (2012)
  [arXiv:1207.0186 [hep-th]].

\bibitem{Armoni:2006ux} 
  A.~Armoni,
  JHEP {\bf 0611}, 009 (2006)
  [hep-th/0608026].
  
\bibitem{Alday:2007hr} 
  L.~F.~Alday and J.~M.~Maldacena,
  JHEP {\bf 0706}, 064 (2007)
  [arXiv:0705.0303 [hep-th]].
  
\bibitem{Alday:2007mf} 
  L.~F.~Alday and J.~M.~Maldacena,
  JHEP {\bf 0711}, 019 (2007)
  [arXiv:0708.0672 [hep-th]].

\bibitem{Boels:2017skl} 
  R.~H.~Boels, T.~Huber and G.~Yang,
  Phys.\ Rev.\ Lett.\  {\bf 119}, 201601 (2017)
  [arXiv:1705.03444 [hep-th]].
  
\bibitem{Boels:2017ftb} 
  R.~H.~Boels, T.~Huber and G.~Yang,
  JHEP {\bf 1801}, 153 (2018)
  [arXiv:1711.08449 [hep-th]].

\bibitem{Dixon:2017nat} 
  L.~J.~Dixon,
  JHEP {\bf 1801}, 075 (2018)
  [arXiv:1712.07274 [hep-th]].

\bibitem{Almelid:2017qju} 
  \O.~Almelid, C.~Duhr, E.~Gardi, A.~McLeod and C.~D.~White,
  JHEP {\bf 1709}, 073 (2017)
  [arXiv:1706.10162 [hep-ph]].
  
\bibitem{Henn:2019swt} 
  J.~M.~Henn, G.~P.~Korchemsky and B.~Mistlberger,
  arXiv:1911.10174 [hep-th].
 
  
\bibitem{Moch:2004pa} 
  S.~Moch, J.~A.~M.~Vermaseren and A.~Vogt,
  Nucl.\ Phys.\ B {\bf 688}, 101 (2004)
  [hep-ph/0403192].
  
\bibitem{Moch:2005id} 
  S.~Moch, J.~A.~M.~Vermaseren and A.~Vogt,
  JHEP {\bf 0508}, 049 (2005)
  [hep-ph/0507039].

\bibitem{Becher:2006mr} 
  T.~Becher, M.~Neubert and B.~D.~Pecjak,
  JHEP {\bf 0701}, 076 (2007)
  [hep-ph/0607228].

\bibitem{tHooft:1972tcz} 
  G.~'t Hooft and M.~J.~G.~Veltman,
  Nucl.\ Phys.\ B {\bf 44}, 189 (1972).

\bibitem{Siegel:1979wq} 
  W.~Siegel,
  Phys.\ Lett.\  {\bf 84B}, 193 (1979).
  
\bibitem{Bern:1991aq} 
  Z.~Bern and D.~A.~Kosower,
  Nucl.\ Phys.\ B {\bf 379}, 451 (1992).

\bibitem{Broggio:2015dga} 
  A.~Broggio, C.~Gnendiger, A.~Signer, D.~St\"ockinger and A.~Visconti,
  JHEP {\bf 1601}, 078 (2016)
  [arXiv:1506.05301 [hep-ph]].



\end{thebibliography}
\end{document}